\documentclass[useAMS]{mn2e}

\usepackage{times}
\usepackage{graphicx}

\begin{document}


\newcommand{\intdv}{$\int$\trstar$\d v$}
\newcommand{\hcop}{HCO$^{+}$}
\newcommand{\kms}{$\,$km$\,$s$^{-1}$}
\newcommand{\mic}{$\mu$m}
\newcommand{\cucm}{cm$^{-3}$}
\newcommand{\sqcm}{cm$^{-2}$}
\newcommand{\degs}{$^{\circ}$}
\newcommand{\msun}{M$_{\odot}$}
\newcommand{\lsun}{L$_{\odot}$}
\newcommand{\tastar}{$T_{\rm A}^{*}$}
\newcommand{\trstar}{$T_{\rm R}^{*}$}
\newcommand{\trad}{$T_{\rm R}$}
\newcommand{\amm}{NH$_{3}$}
\newcommand{\too}{$\rightarrow$}
\newcommand{\ceo}{C$^{18}$O}
\newcommand{\cso}{C$^{17}$O}
\newcommand{\thco}{$^{13}$CO}
\newcommand{\twco}{$^{12}$CO}
\newcommand{\ci}{C\,{\textsc i}}
\newcommand{\lkha}{LkH$\alpha$101}
\newcommand{\HII}{H\,{\sc ii}}

\title[High-frequency radio observations of massive YSOs]{A high-frequency radio
  continuum study of massive young stellar objects}

\author[A.G. Gibb \& M.G. Hoare] 
{A.G. Gibb$^{1}$, M.G. Hoare$^{2}$\\
$^1$Department of Physics and Astronomy, University of British
Columbia, 6224 Agricultural Road, Vancouver, B.C. V6T 1Z1, Canada\\
$^2$School of Physics and Astronomy, University of Leeds, Leeds,
West Yorkshire, LS2 9JT
}

\date{Accepted 2007 May 25. Received 2007 May 14; in original form 2007 March 16}

\maketitle

\begin{abstract}
  We present high-resolution observations made with the Very Large
  Array (VLA) in its A configuration at frequencies between 5 and 43
  GHz of a sample of five massive young stellar objects (YSOs): \lkha,
  NGC2024-IRS2, S106-IR, W75N and S140-IRS1. The resolution varied
  from 0.04 arcsec (at 43 GHz) to 0.5 arcsec (at 5 GHz), corresponding
  to a linear resolution as high as 17 AU for our nearest source. A
  MERLIN observation of S106-IR at 23 GHz with 0.03-arcsec resolution
  is also presented.
  S106-IR and S140-IRS1 are elongated at 43 GHz perpendicular to their
  large scale bipolar outflows. This confirms the equatorial wind
  picture for these sources seen previously in MERLIN 5 GHz
  observations. The other sources are marginally resolved at 43 GHz.
  The spectral indices we derive for the sources in our sample range
  from +0.2 to +0.8, generally consistent with ionized stellar
  winds. We have modelled our sources as uniform, isothermal spherical
  winds, with \lkha\ and NGC\,2024-IRS2 yielding the best
  fits. However, in all cases our fits give wind temperatures of only
  2000 to 5000\,K, much less than the effective temperatures of
  main-sequence stars of the same luminosity, a result which
  is likely due to the clumpy nature of the winds.
\end{abstract}


\section{Introduction}

Like their low-mass counterparts, young massive stars drive molecular
outflows. The outflows from low-mass young stellar objects (YSOs)
appear to be driven by highly-collimated jets emanating from the
central star-disc region (e.g., Reipurth et al.\ 2002; Burrows et al.\
1996; Ouyed, Clarke \& Pudritz 2003). The situation for massive YSOs
is less clear: only in a small number of cases have jets been
detected, and those detections have all been at centimetre wavelengths
(e.g., HH\,80--81 -- Mart\'{\i} et al.\ 1993; Cep\,A-HW2 --
Rodr\'{\i}guez et al.\ 1994; W3-H$_2$O -- Wilner, Reid \& Menten 1999;
and G35.2$-$0.7N -- Gibb et al.\ 2003). Surprisingly, {\em
equatorially}-elongated radio emission has been detected towards three
sources: S\,106 (Hoare et al.\ 1994), S\,140-IRS1 (Hoare 2006) and
more recently Orion-I (Reid et al.\ 2007). Hoare (2002) summarizes the
radio properties of massive YSOs.

The primary mechanism for the radio continuum emission from massive
YSOs is thermal free-free emission from ionized gas. This may either
be in the form of a photo-ionized \HII\ region or stellar wind. An
infinite, constant velocity ionized wind has a spectral index of +0.6
(Wright \& Barlow 1975; Panagia \& Felli 1975), while that for an
\HII\ region varies from $-$0.1 (optically thin) to +2 (optically
thick). However, if the wind is finite in extent (i.e.\ the ionized
gas recombines to form a neutral wind), then the spectral index can
increase as the wind begins to take on the appearance of an optically
thick \HII\ region (Simon et al.\ 1983; Moran et al.\ 1983). For
collimated flows, the spectral index (for unresolved sources) can take
on values between 0.25 and 1.1, although a conical flow yields the
same spectral index as expected for a spherical wind (Reynolds 1986).

Infrared recombination line studies have shown that the winds from
massive YSOs can reach speeds of several hundred \kms\ (Drew, Bunn \&
Hoare 1993; Bunn, Hoare \& Drew 1995). However, these spectral studies
are not able to resolve the geometry of the high-velocity gas. For
this we must turn to the radio to achieve the high spatial resolution
necessary to probe the small-scale structure of the ionized gas.

Although theoretically attractive, it remains unclear whether a
circumstellar disc forms during the collapse of a massive cloud core
and if so whether it survives the pre-main sequence evolution of a
massive YSO given the detrimental impact of the wind and ionizing
radiation (e.g.\ Richling \& Yorke 1997). The observation of jet-like
outflows suggests that discs do indeed form in some cases, while
recent high-resolution submillimetre imaging has yielded good
observational evidence for a disc around a massive YSO (Patel et al.\
2005).

In this paper we present the results of a high-resolution,
multi-wavelength radio survey of a sample of five massive YSOs (Table
\ref{sources}), none of which has evidence for a jet. Furthermore,
none of these have been previously observed at 43 GHz, nor with such
high angular resolution. Previous 43 GHz studies of high-mass YSOs
have been published by Menten \& van der Tak (2004), van der Tak \&
Menten (2005) and van der Tak, Tuthill, \& Danchi (2005), and these
are discussed later. Menten \& Reid (1995) and Rodr\'{\i}guez et al.\
(2005) present 43-GHz observations of BN and Orion-I, although these
papers do not discuss the physical properties derived from these
data. The high-frequency radio continuum spectra of these sources are
discussed by Plambeck et al.\ (1995). The aims of the current study
were to explore the high-frequency characteristics via the spectral
index (for clues to emission mechanisms) and small-scale geometry of
the ionized winds from massive YSOs. This paper presents the results
of these observations, and provides new information on the
geometry of the ionized gas close to these massive YSOs.

\begin{table}
\centering
\caption{Source properties. Coordinates are of the phase centres used
  in the observations. The notation $a(b)$ represents $a\times 10^b$.}
\begin{tabular}{lcccc}
Source &    RA    &    Dec    &  $d$  &   $L$       \\
       &  (B1950) &  (B1950)  & (kpc) & (L$_\odot$) \\
\hline
S\,106-IR         & 20:25:33.8 &   +37:12:49.98 & 0.6  & 2.0(4) \\
S\,140-IRS1       & 22:17:41.1 &   +63:03:41.58 & 0.9  & 5.0(3) \\
W75N              & 20:36:50.0 &   +42:26:57.13 & 2.0  & 4.4(4) \\
NGC\,2024-IRS2    & 05:39:14.3 & $-$01:55:55.00 & 0.42 & 1.0(4) \\
Lk\,H$\alpha$101  & 04:26:57.2 &   +35:09:54.95 & 0.7  & 4.0(4) \\
\end{tabular}
\label{sources}
\end{table}

\begin{table}
\centering
\caption{Coordinates of detected sources.}
\begin{tabular}{lccc}
Source & RA(B1950) & Dec(B1950) & ID \\
\hline
S\,106-IR         & 20:25:33.823  &  ~+37:12:50.04 & IR/IRS4 \\
S\,140-IRS1       & 22:17:41.085  &  ~+63:03:41.58 & IRS1 \\
                  & 22:17:41.051  &  ~+63:03:59.41 & IRS2N \\
W75N              & 20:36:50.005  &  ~+42:26:58.49 & VLA1/Ba \\
                  & 20:36:50.042  &  ~+42:26:57.13 & VLA3/Bb \\
                  & 20:36:50.084  &  ~+42:26:55.46 & Bc \\
NGC\,2024-IRS2    & 05:39:14.301  & $-$01:55:55.00 & IRS2 \\
Lk\,H$\alpha$101  & 04:26:57.235  &  ~+35:09:54.95 & \lkha \\
\end{tabular}
\label{peaks}
\end{table}

\begin{table*}
\centering
\caption{Calibrator properties. Amplitude calibrators are denoted by
   an `A' in the Calibrator Type column. The Notes column denotes how
   the flux was obtained: `D' -- derived within AIPS, `A' -- assumed,
   `B' -- bootstrapped from amplitude calibrator. Fluxes for 0134+329
   (at 5 and 8.5 GHz) and 0316+329 (at 22.5 GHz) were derived using
   standard VLA calibration procedures within AIPS. Fluxes for the
   phase calibrators (denoted by a `P' in the Calibrator Type column)
   are bootstrapped values derived from the amplitude
   calibrators. Note that the uncertainties in the bootstrapped 43-GHz
   fluxes does not include the uncertainty in the absolute
   calibration.}
\begin{tabular}{lcccccc}
Calibrator &  \multicolumn{4}{c}{Flux density (Jy)}    & Calibrator  & Notes \\
           &  5 GHz & 8.5 GHz & 22.5 GHz & 43 GHz & Type & \\
\hline
0134+329  & 5.51$\pm$0.03 & 3.22$\pm$0.01 & 1.12$\pm$0.01   & --   & A & D \\
0710+439  &   -- &  --  & --   & 0.3$\pm$0.1 & A & A \\
\hline
0440+345  & 1.00$\pm$0.01 & 0.83$\pm$0.02 & 0.34$\pm$0.02 &  --   & P & B \\
0441+341  &  --  &  --  & --   &  0.52$\pm$0.01 & P & B\\
0539$-$057& 0.95$\pm$0.02 & 1.00$\pm$0.03 & 0.76$\pm$0.04 &  0.46$\pm$0.02 & P & B\\
2005+403  &  --  &  --  & --   &  1.09$\pm$0.03 & P & B\\
2023+336  & 2.80$\pm$0.03 & 2.60$\pm$0.06 & 1.72$\pm$0.05 &  2.56$\pm$0.08 & P & B\\
2229+695  & 0.37$\pm$0.01 & 0.40$\pm$0.01 & 0.39$\pm$0.02 &  0.31$\pm$0.01 & P & B\\
\hline
\end{tabular}
\label{cals}
\end{table*}

\begin{table*}
\centering
\caption{Observational parameters. The noise level (1$\sigma$ in
mJy\,beam$^{-1}$) and beam dimensions (in arcsec) are those of the
maps presented in this paper. The number in parentheses after the beam
dimensions is the beam position angle, measured in degrees east of
north. \label{obsparms}}
\begin{tabular}{lcccccccc}
 & \multicolumn{2}{c}{5 GHz} & \multicolumn{2}{c}{8.5 GHz} &
       \multicolumn{2}{c}{22.5 GHz} & \multicolumn{2}{c}{43 GHz}\\
Source       &  Noise  & Beam         & Noise   & Beam         &       Noise     & Beam         & Noise   & Beam         \\ 
\hline
S\,106-IR         &   --- &               ---        & 0.064 & 0.26$\times$0.19 ($-$23) & 0.25 & 0.100$\times$0.069 ($+$6) & 0.35 & 0.040$\times$0.031 ($-$69) \\ 
S\,140-IRS1       & 0.041 & 0.55$\times$0.31 ($-$27) & 0.034 & 0.32$\times$0.18 ($-$23) & 0.28 & 0.120$\times$0.071 ($-$7) & 0.36 & 0.042$\times$0.031 ($-$45) \\
W75N              & 0.048 & 0.50$\times$0.33 ($-$35) & 0.034 & 0.29$\times$0.19 ($-$32) & 0.32 & 0.106$\times$0.077 ($-$4) & 0.30 & 0.041$\times$0.031 ($-$63) \\
NGC\,2024-IRS2    & 0.200 & 0.67$\times$0.51 ($-$34) & 0.091 & 0.43$\times$0.30 ($-$43) & 0.47 & 0.138$\times$0.102 ($+$83)& 0.40 & 0.063$\times$0.039 ($-$42) \\ 
Lk\,H$\alpha$101  & 0.043 & 0.55$\times$0.36 ($+$61) & 0.055 & 0.33$\times$0.22 ($+$67) & 0.30 & 0.129$\times$0.069 ($+$72)& 0.31 & 0.041$\times$0.034 ($-$89) \\
\end{tabular}
\end{table*}

\section{Observations}
\label{observations}
The observations were made using the Very Large Array (VLA) of the
National Radio Astronomy Observatory in its most extended (A)
configuration. The data were recorded on 1996 November 1. Each of the
sources listed in Table \ref{sources} was observed at C (4.86 GHz), X
(8.46 GHz), K (22.46 GHz) and Q (43.49 GHz) band.  At the time of
observation, the VLA had 43-GHz receivers on 13 antennas, leaving 14
to be used at other frequencies (although only 12 were useable). Our
observing procedure was to observe the source for the full on-source
time with the Q-band subarray and split the observing time for the
non-Q band subarray between the other three frequencies. The typical
on-source time was 9 minutes at 5 GHz, 9 minutes at 8.5 GHz, 7 to 15
minutes at 22.5 GHz and 30 minutes at 43 GHz.

The minimum and maximum telescope separations at each frequency was
1.6 to 25.6 km (C, X and K), and 1.0 to 34.6 km (Q).  The resolution
at each of these frequencies ranges from 0.4 arcsec (C band) to 35
milliarcsec (Q band). At the distance of our closest source
(NGC\,2024-IRS2) the resolution at 43 GHz corresponds to $\sim$17 AU,
and 85 AU for our most distant source (W75N).

Absolute amplitude calibration was carried out using 0134+329 (C, X
and K band) and 0710+439 (Q band). Calibrator details are given in
Table \ref{cals}, where the names are given in B1950 form. A model was
used for 0134+329 at K band. Although 0134+329 is a VLA standard
calibrator, it could not be used at 43 GHz due to limitations in the
number of antennas near the centre of the array in this
subarray. Therefore, the flux of 0710+439 has been assumed based on
previous observations. A correction for atmospheric opacity was made
at 43 GHz, as tipping scans yielded a zenith optical depth of 0.1. The
estimated calibration uncertainty is less than one per cent at 5, 8.5
and 22.5 GHz. At 43 GHz the absolute uncertainty is much higher at
$\sim$30 per cent, although the relative uncertainty is small,
typically less than 5 per cent. 

Nearby quasars were used for phase referencing (denoted by `P' in
Table \ref{cals}). The `fast-switching' technique was employed at 43
GHz, in order to limit decorrelation due to atmospheric phase
fluctuations, with a cycle time of approximately 2 minutes (Carilli \&
Holdaway 1999). The bootstrapped flux densities of the phase
calibrators are generally self-consistent, although the K-band flux
for 2023+336 appears to be anomalously low. 

The data were edited and calibrated using the AIPS package, and
reduced and analyzed using AIPS and MIRIAD (Sault et al.\
1995). Images were constructed using a variety of weighting schemes by
varying the Briggs robust parameter. In general, a neutral value of 0
gave the optimum balance between resolution and sensitivity to any
extended emission. Table \ref{obsparms} lists the beam parameters and
sensitivity for each source at each wavelength.

\subsection{Additional observations}

We also observed S106-IR at 23 GHz with MERLIN on 1996 May 5. 3C273
was used for pointing and absolute calibration, assuming a flux
density of 20 Jy. 2023+335 was used for phase calibration. In order to
obtain a reasonable image, a 15 M$\lambda$ taper was applied to the
data to give a beam of 0.031$\times$0.028 arcsec$^2$ and a noise level
of 0.95 mJy\,beam$^{-1}$. Only four of the MERLIN antennas were used
so the image fidelity is poor. However, the general size and shape of
the emission are robust.

S106-IR and NGC2024-IRS2 were also observed at 86 and 106 GHz with the
Plateau de Bure interferometer (PdBI) on several dates during 1996
September and October with the antennas in the C and D
configurations. The data were calibrated using the CLIC software
package. NRAO\,150 was used for bandpass calibration, 0458$-$020 and
2013+370 for phase calibration. The amplitude calibration was derived
by a bootstrap method, solving iteratively for the antenna gains. The
beamsize was $\sim$4--4.5 arcsec and noise level was
10--20\,mJy\,beam$^{-1}$. Fluxes were derived from Gaussian fits to
the visibility data. S106-IR was not detected (although extended
emission was detected; see Bieging 1984); NGC2024-IRS2 was
detected as a point source with a flux density of 60/110 mJy at 86/106
GHz respectively.

\section{Results \& Analysis}

In this section we introduce each source, present images at each
frequency and calculate spectral indices. Visibility amplitude plots
are shown in section \ref{modelling}. Source positions were derived
from 2-D gaussian fits at each wavelength and found to be consistent
at all wavelengths to within 0.04 to 0.06 arcsec, in good agreement
with the expected accuracy stated in the VLA Observational Status
Summary. The position of each detected source is listed in Table
\ref{peaks}.

Flux estimates for each source were derived in two ways. The first was
to sum the emission within a region encompassing the source, and the
second was to carry out a two-dimensional gaussian fit. In most cases,
the sources were well-fit by gaussians and the dimensions listed for
each source are the deconvolved major and minor axes derived from
these fits (Tables \ref{fits} and \ref{sizes}). Any significant
deviations from a gaussian fit are discussed in the text. Gaussian
fits to the visibility data gave poor results, significantly
over-estimating the total flux and were therefore not used in our
analysis. The uncertainties in the peak and total fluxes given in
Table \ref{fits} represent the mean residual between the source and a
gaussian fit, and do not include contributions from the uncertainty in
the absolute calibration. Brightness temperatures were derived from
the peak fluxes using the beam sizes given in Table
\ref{obsparms}. These will be lower limits to the actual temperature
of the gas due to low optical depth and/or incomplete beam filling.

The centimetre spectral energy distribution for each source was fitted
between 5 and 43 GHz to derive a spectral index (also listed in Table
\ref{fits}). In addition we calculated the index for the angular size
as a function of frequency, which we denote by $\zeta$ (and list in
Table \ref{sizes}). We also plot 3-mm fluxes on the spectral energy
distribution for comparison.

\subsection{S106-IR}

\begin{figure*}
\centering
\includegraphics[width=16.25cm]{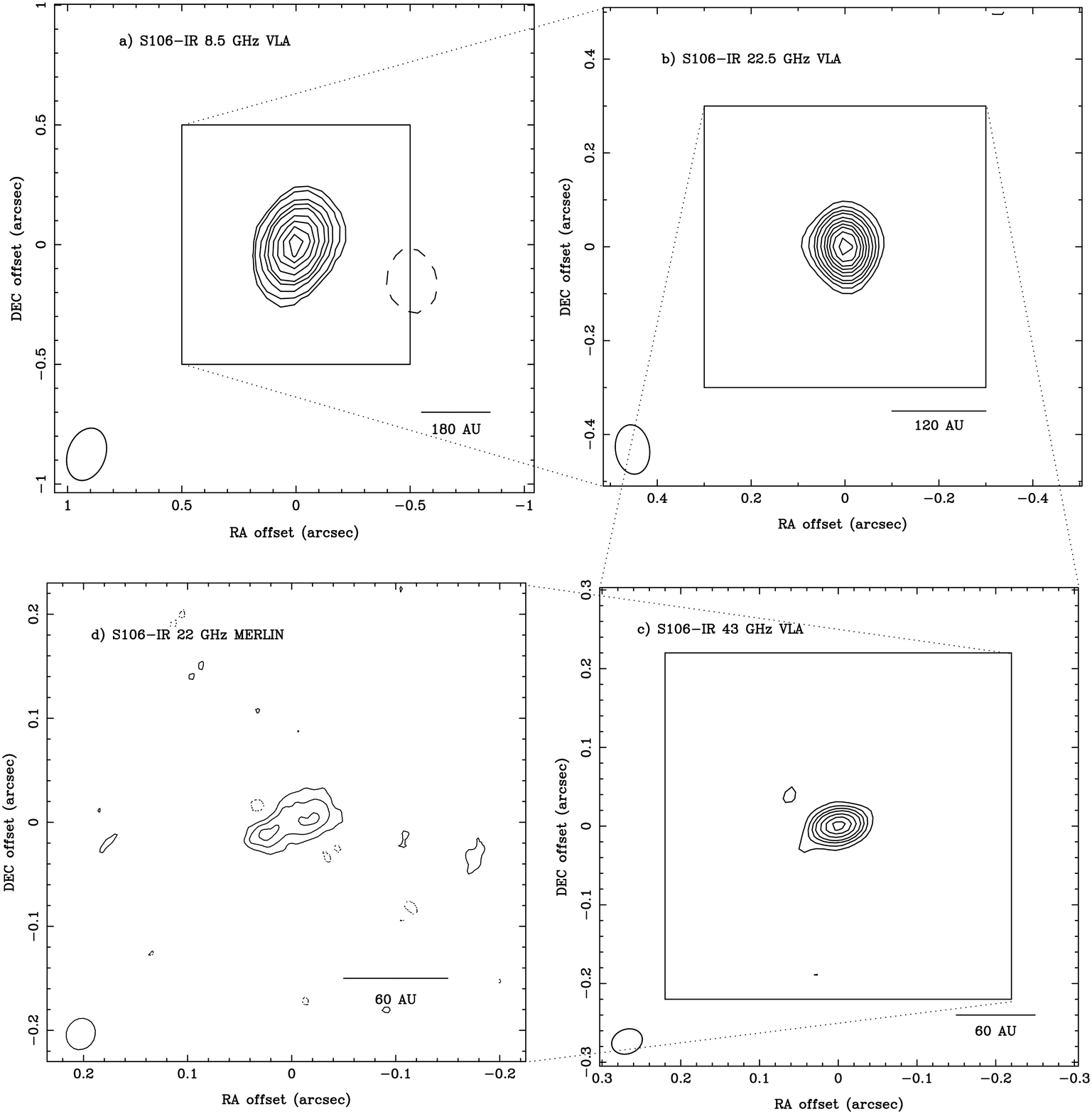}
\caption{Images of S106-IR at each frequency, clockwise from top left:
  a) 8.5 GHz, b) 22.5 GHz, c) 43 GHz and d) 22 GHz MERLIN data.
  Contours are at a) $-$3, 3, 5, 10, 15, 20, 30, 40, 50,
  60$\times$0.064 mJy\,beam$^{-1}$; b) $-$3, 3, 5, 7, 9, 12, 15, 18,
  21, 25, 30, 35$\times$0.25 mJy\,beam$^{-1}$; c) $-$3, 3, 5, 7, 9,
  12, 15, 18$\times$0.35 mJy\,beam$^{-1}$; and d) $-$1, 1, 2,
  3$\times$1.5 mJy\,beam$^{-1}$ respectively. The large box in a), b)
  and c) represents the size of the next image. The open ellipse in
  the bottom left corner represents the beam, and a scale bar is shown
  as a horizontal line assuming the distance given in Table
  \ref{sources}. The coordinates for the MERLIN image given by
  Hoare \& Muxlow (1996) were corrected for the different epoch used
  at the VLA.}
\label{s106}
\end{figure*}

S106 is a well-known optical and radio bipolar \HII\ region (Bally,
Snell \& Predmore 1983; Bally et al.\ 1998). Bally et al.\ (1983)
detected a compact radio source at the centre of the \HII\ region
coincident with the infrared source S106-IR (also known as IRS4).
Mid-infrared imaging by Smith et al.\ (2001) show that S106-IR is the
sole exciting source of the large-scale ionized nebula. Infrared
spectroscopy by Drew, Bunn \& Hoare (1993) revealed the presence of a
highly supersonic, ionized wind with a terminal velocity exceeding 340
\kms\ and inferred that enhanced mass-loss must be occurring along the
equatorial direction.  High-resolution radio observations at 5\,GHz
with MERLIN revealed the ionized gas takes on an equatorial
configuration in the radio with a position angle of $-$61$\pm$2
degrees (Hoare et al.\ 1994). The X-ray results of Giardino et al.\
(2004) are consistent with shocks in a radiatively-driven stellar
wind. Near-infrared speckle observations by Alvarez et al.\ (2004)
show only a single unresolved source.

We observed S106-IR at 8.5, 22.5 and 43 GHz (Fig. \ref{s106}). In
addition we present MERLIN 23-GHz data for comparison in
Fig. \ref{s106}d (see also Hoare \& Muxlow 1996). Contamination from
the large-scale \HII\ region was a significant problem at 8.5 GHz; we
only include data on baselines greater than 200 k$\lambda$ in
Fig. \ref{s106}a. Compact emission was detected only from S106-IR; the
nearby protostellar source S106-FIR was not detected to a 3-$\sigma$
limit of 0.15 mJy at 8.5 GHz and 1.1 mJy at 43 GHz. 

S106-IR is marginally resolved at 22.5 GHz, and is clearly resolved
(along the long axis) at 43 GHz showing an elongated geometry a
direction perpendicular to the large-scale \HII\ region. The 5-GHz
MERLIN observations of Hoare et al.\ (1994) revealed a source
elongated perpendicular to the direction of the large-scale bipolar
\HII\ region. Our results show that the emission at 43 GHz has a
similar geometry, reinforcing the interpretation of Hoare et al.\
(1994) that the ionized gas is confined to an equatorial
configuration. The MERLIN map at 23 GHz (Fig.\ \ref{s106}d) coupled
with the fact that there is no known outflow with this position angle
confirms the equatorial geometry beyond doubt.

The gaussian fit to the 22.5-GHz VLA image in Fig. \ref{s106}b (made
with a robust parameter of $-$2) yields a point source upon
deconvolution. However, we note that there is a slight east-west
elongation in the lower contours for this image. The fit to an image
made with a robust parameter of +2 yields an east-west deconvolved
source geometry: 0.057$\times$0.013 arcsec$^2$ at a position angle of
$-$87 degrees. This is in reasonable agreement with the fit to the
MERLIN 23-GHz data which gives 0.074$\times$0.03 arcsec$^2$ at a
position angle of $-$63 degrees. The VLA 43-GHz data are in excellent
agreement with the MERLIN data, showing a position angle of $-$61
degrees.

The VLA 22.5-GHz flux measurement is in good agreement with MERLIN
measurement within the uncertainties. Both measurements are slightly
lower than the 17 mJy measured by Felli et al.\ (1984). We see no
evidence of the double-peaked appearance in our VLA observations
compared with the MERLIN data but our highest VLA resolution
($\sim$0.035 arcsec at 43 GHz) is barely sufficient to resolve these
maxima.

We derive a spectral index of +0.65$\pm$0.11 from the VLA data, in
good agreement with the value of 0.7 derived by Bally et al.\
(1983). However, we note that our fluxes are consistently lower by
$\sim$60 per cent, which is probably a result of our observations
being less sensitive to extended emission than previous. The
angular-size as a function of frequency for the VLA data is flatter
than expected for a uniform spherical wind, but only two points
contribute to this fit. The higher-resolution MERLIN data (although
also only two points) give a value for $\zeta$ of $-$0.59, very close
to the value predicted for a uniform wind. Based on the MERLIN results
of Hoare et al.\ (1994) and the data presented here, the predicted
source diameter at 43 GHz is $\sim$0.05 arcsec, a little larger than
our derived value of 0.031 arcsec.

\begin{figure*}
\centering
\includegraphics[width=16.25cm]{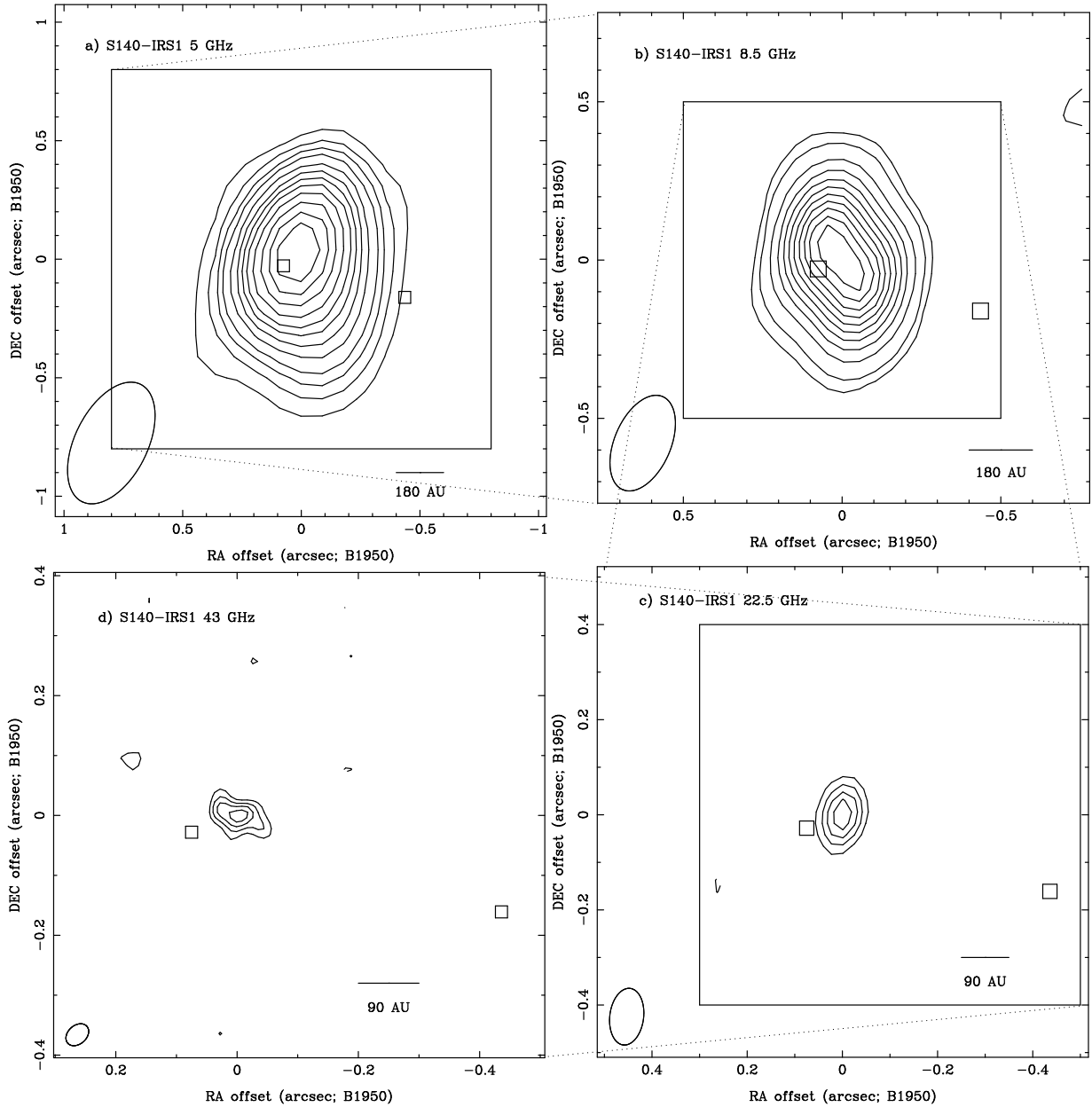}
\caption{Images of S140-IRS1 at each frequency, clockwise starting
  from top left: a) 5 GHz, b) 8.5 GHz, c) 22.5 GHz and d) 43 GHz. Open
  squares are water masers from Tofani et al.\ (1995). Contours are
  at: a) $-$3, 3, 5, 7, 9, 12, 15, 18, 21, 25, 30, 35, 40$\times$0.041
  mJy\,beam$^{-1}$; b) $-$3, 3, 5, 10, 15, 20, 25, 30, 35, 40, 45, 50,
  55, 60, 65, 70, 75, 80$\times$0.034 mJy\,beam$^{-1}$; c) $-$3, 3, 5,
  7, 9, 12, 15$\times$0.28 mJy\,beam$^{-1}$ and d) $-$3, 3, 5, 7, 9,
  12$\times$0.36 mJy\,beam$^{-1}$ respectively. The large box in a),
  b) and c) represents the size of the next image. The open ellipse in
  the bottom left corner represents the beam, and a scale bar is shown
  as a horizontal line assuming the distance given in Table
  \ref{sources}. }
\label{s140}
\end{figure*}

The brightness temperature increases with frequency, indicating that
the higher-frequency observations are tracing hotter gas close to the
star. However, the brightness temperatures derived from the VLA data
are all considerably lower than either the likely equivalent
main-sequence effective temperature of 22\,000\,K or even half that
value (see Drew 1989). Comparing these values with those derived from
the MERLIN data (18\,000 and 13\,500\,K at 5 and 23 GHz respectively)
suggests that the emission is clumpy and has structure on scales
smaller than the VLA beam.

\subsection{S140-IRS1}

S140-IRS1 is the brightest of a group of infrared sources and has long
been known to be associated with radio emission and a bipolar CO
outflow (Simon et al.\ 1983; Schwartz 1989; Minchin, White \& Padman
1993b). A compact near-infrared reflection nebula is observed in the
direction of the blue lobe of the CO outflow (Hoare, Glindemann \&
Richichi 1996; Schertl et al.\ 2000; Alvarez et al.\ 2004). IRS1 is
also associated with water masers (Tofani et al.\ 1995) and its radio
emission is elongated perpendicular to the reflection nebula and
outflow directions (Hoare 2006). Thus S\,140-IRS1 is a second example
of a massive YSO with an equatorial wind.

S140-IRS1 was observed at all four frequencies in our survey (Fig
\ref{s140}). IRS1 is clearly detected and is resolved at 5, 8.5 and 43
GHz; only at 22.5 GHz does it appear to be a point source, probably
due to a lack of sensitivity. The 43-GHz emission is especially
interesting as we clearly resolve structure in the ionized gas on
scales of 40 AU. The emission is elongated in a similar direction to
that seen previously at lower frequencies (Schwartz 1989; Tofani et
al.\ 1995; Hoare 2006). The position angle (PA) of the extended arms
is $\sim$60--65 degrees east of north, compared with 35 degrees for
the 5 and 8.5-GHz emission. A line drawn through the tips of the
extended emission defines a PA of $\sim$45 degrees. The MERLIN 5-GHz
observations shown by Hoare (2006) have a very similar PA of $\sim$45
degrees, although that emission is more extended than our 43-GHz
data. However, unlike the 5-GHz MERLIN data, which shows linearly
extended emission, the 43-GHz image displays a clockwise twist in the
lower contours, which gives the appearance of anti-clockwise rotation
(Hoare 2006). Furthermore, the proper motion of the ionized gas is
consistent with the rotation interpretation (Hoare 2006). There is good
agreement between our 8.5-GHz image (Fig. \ref{s140}b) and that of
Tofani et al.\ (1995), showing the same twist in the lower contours.

The spectral index of IRS1 is well-constrained and consistent with a
constant velocity ionized wind ($\alpha$ =
0.61$\pm$0.03). Furthermore, the deconvolved size of the emission from
IRS1 falls off in a manner close to that expected for a uniform
spherical wind. Previous estimates of the spectral index give a
slightly higher value of 0.8 (Schwartz 1989; Tofani et al.\ 1995) for
the central source and a smaller value for the extended emission
(interpreted as a jet by Schwartz 1989). The brightness temperatures
at 5 and 8.5 GHz are rather similar at a little less than 600\,K
(Table \ref{fits}). They increase at 22.5 and 43 GHz but remain below
2000\,K. Extrapolating to higher frequencies shows that the wind from
IRS1 contributes a significant fraction of the flux at 3 mm
(Fig. \ref{sed}).

The 5-GHz MERLIN data of Hoare \& Muxlow (1996) and Hoare (2006)
reveal a more elongated source which is somewhat larger than predicted
by the value for $\zeta$ given in Table \ref{sizes}. This may be a
result of our observations not detecting all of the emission. An
intriguing fact is that the 43-GHz data has a different position angle
to the MERLIN data, although it must be noted that the extent of the
43-GHz emission lies within the central part of the MERLIN image.

We also detected IRS2N at 5 and 8.5 GHz, with a spectral index of
$\sim$0.77, in reasonable agreement with previous determination by
Schwartz (1989), at least for the 1.6 and 5 GHz data. It is likely
that IRS2N is an HII region, with a turnover frequency below 15 GHz
(c.f.\ Schwartz 1989), and is optically thin and resolved at 22.5 and
43 GHz.

IRS2S was not detected above a 3-$\sigma$ level of 0.1
mJy\,beam$^{-1}$ at 8.5 GHz, despite its clear detection with a flux
density of $\sim$1 mJy at 5 and 8.4 GHz by Schwartz (1989) and Tofani
et al.\ (1995) respectively. IRS3 was not detected to the same levels.

IRS2S was detected by Tofani et al.\ (1995) at 8.5 GHz with a flux
density of 1.2 mJy\,beam$^{-1}$ in 1992 November, compared with our
3$\sigma$ upper limit of 0.1 mJy\,beam$^{-1}$ in 1996 November, a
reduction of more than an order of magnitude over a 4-year
timescale. Tofani et al.\ (1995) derived a spectral index for IRS2S of
0.6 (between 1.5 and 8.5 GHz), as expected for a constant velocity
wind. Therefore, IRS2S is an embedded massive YSO driving an ionized
wind. Our non-detection may be due to resolving out of the emission,
but we note that the observations by Tofani et al.\ (1995) also used
the VLA in A-configuration. Therefore it is more likely that IRS2S is
variable at centimetre wavelengths.

The `bullet' source (VLA4 in Simon et al. 1983) was not detected in
any of our images despite having sufficient sensitivity to detect
it. It is therefore likely (and perhaps suggested by previous images)
that this source is extended and is resolved out by our observations.

No radio emission was detected towards the location of the
submillimetre source SMM1 (Minchin, Ward-Thompson \& White 1993a) to a
3-$\sigma$ limit of 1.1 mJy at 43 GHz (0.1 mJy\,beam$^{-1}$ at 8.5
GHz).

\subsection{W75N}

\begin{figure*}
\centering
\includegraphics[width=16.5cm]{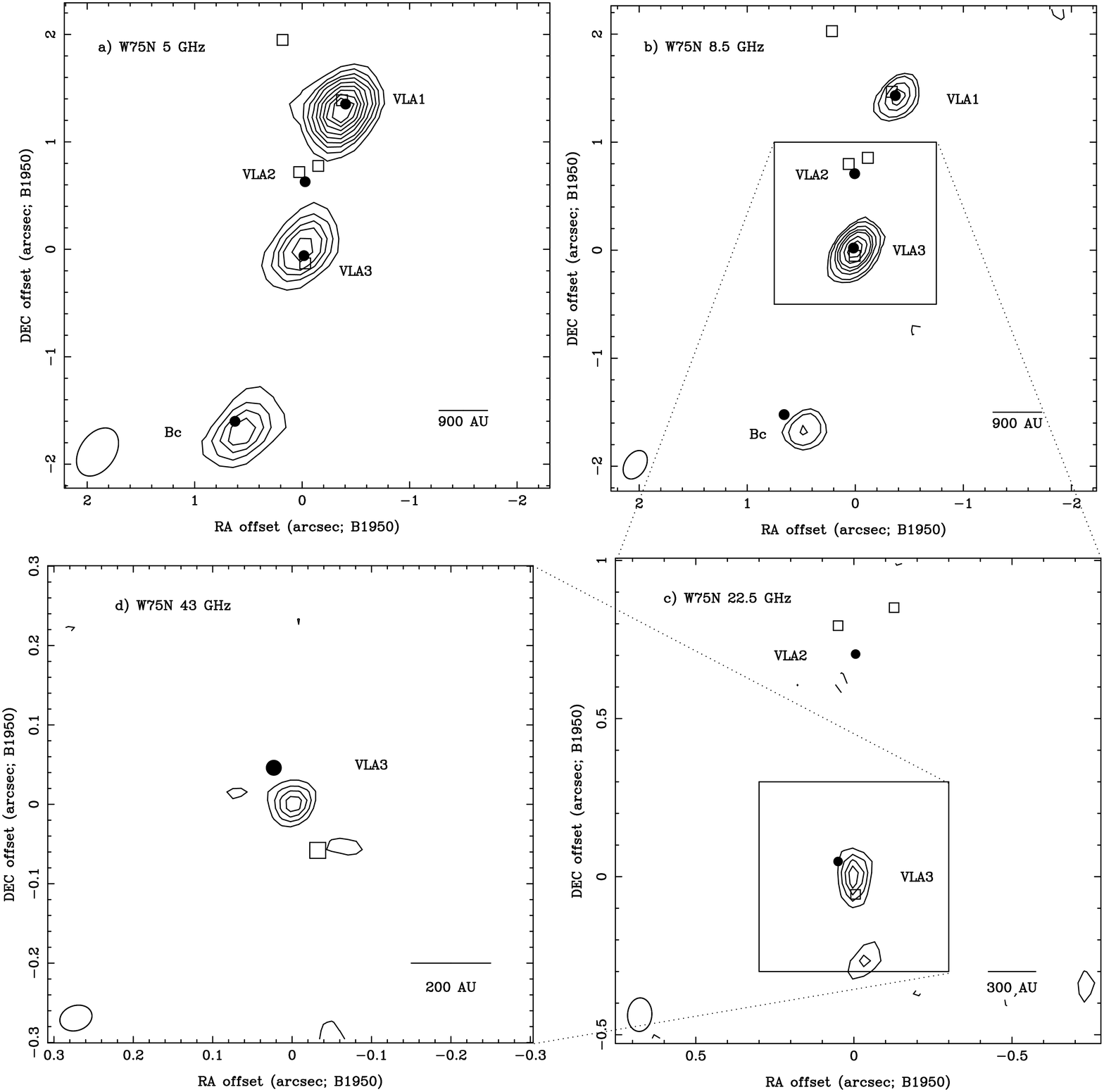}
\caption{Images of W75N at each frequency, clockwise starting from top
  left: a) 5 GHz, b) 8.5 GHz, c) 22.5 GHz and d) 43 GHz. Open squares
  are water masers from Hunter et al.\ (1994) and filled circles are
  radio positions from Hunter et al.\ and Torrelles et al.\
  (1997). Contours are at: a) $-$3, 3, 5, 7, 9, 12, 15, 18, 21,
  24$\times$0.048 mJy\,beam$^{-1}$; b) $-$3, 3, 5, 10, 15, 20, 30, 40,
  40$\times$0.034 mJy\,beam$^{-1}$; c) $-$3, 3, 5, 7, 9$\times$0.32
  mJy\,beam$^{-1}$ and d) $-$3, 3, 5, 7, 9$\times$0.3 mJy\,beam$^{-1}$
  respectively. The large box in the b) and c) represents the size of
  the next image. The open ellipse in the bottom left corner
  represents the beam, and a scale bar is shown as a horizontal line
  assuming the distance given in Table \ref{sources}. }
\label{w75n}
\end{figure*}

W75N contains a small cluster of radio sources associated with a group
of water masers (Hunter et al.\ 1994). A large-scale CO outflow is
observed orientated in a predominantly east-west direction (Davis et
al.\ 1998), although higher-resolution studies have shown that the
flow is composed of several overlapping outflows from a small cluster
of YSOs (Shepherd, Testi, \& Stark 2003). The main group of radio
sources lie within a region approximately 3 arcsec in diameter and lie
in a north-south arrangement. Hunter et al.\ (1994) resolved 3
sources, while observations at 22 GHz resolved four (VLA1--4:
Torrelles et al.\ 1997). VLA1 has a jet-like appearance with the same
position angle as the larger scale CO outflow seen by Davis et al.\
(1998). Shepherd (2001) detected a compact dust source coincident with
the group of radio sources peaking on VLA3 (= W75N\,Bb in the
nomenclature of Hunter et al.\ 1994).

We imaged W75N at all four wavelengths (Fig. \ref{w75n}). At 5 and 8.5
GHz we detect the three sources which make up W75N\,B (Hunter et al.\
1994). At 22.5 and 43 GHz we only detect the most deeply embedded
source VLA3 (W75N\,Bb). The 43-GHz flux density of VLA3 may be as much
as 50 per cent higher since images made with natural weighting recover
a larger flux. However, we note that Shepherd, Kurtz \& Testi (2004)
report a similar value to that we measure from our image (5.7 mJy
compared with 5.2 mJy). From a gaussian fit, VLA3 appears to be
resolved at 43 GHz: its appearance in Fig. \ref{w75n}d is circular
despite the elongated beam. It is interesting to note that the
position angle of the 43-GHz emission (due north) is almost
perpendicular to the proposed outflow direction (PA of 101 deg) from
Shepherd et al.\ (2003). However, it is also vastly different to the
position angle of the 22-GHz emission imaged by Torrelles et al.\
(1997), although our source size is in good agreement with
theirs. Clearly, further high-resolution observations are desirable to
constrain the small-scale geometry of this source.

We do not detect the source VLA2 at any wavelength, probably for a
combination of it being too faint (its flux density at 22.5 GHz is
only $\sim$4 times our noise level: Torrelles et al.\ 1997) and too
extended for our observations (Shepherd et al.\ 2004 detect it with a
larger beam at 43 GHz with a similar sensitivity to ours). VLA1 is too
weak for us to detect at 22.5 GHz: its peak flux density is roughly at
our 4-$\sigma$ level (Torrelles et al.\ 1997).

\begin{figure*}
\centering
\vspace{0.5cm}
\includegraphics[width=17cm]{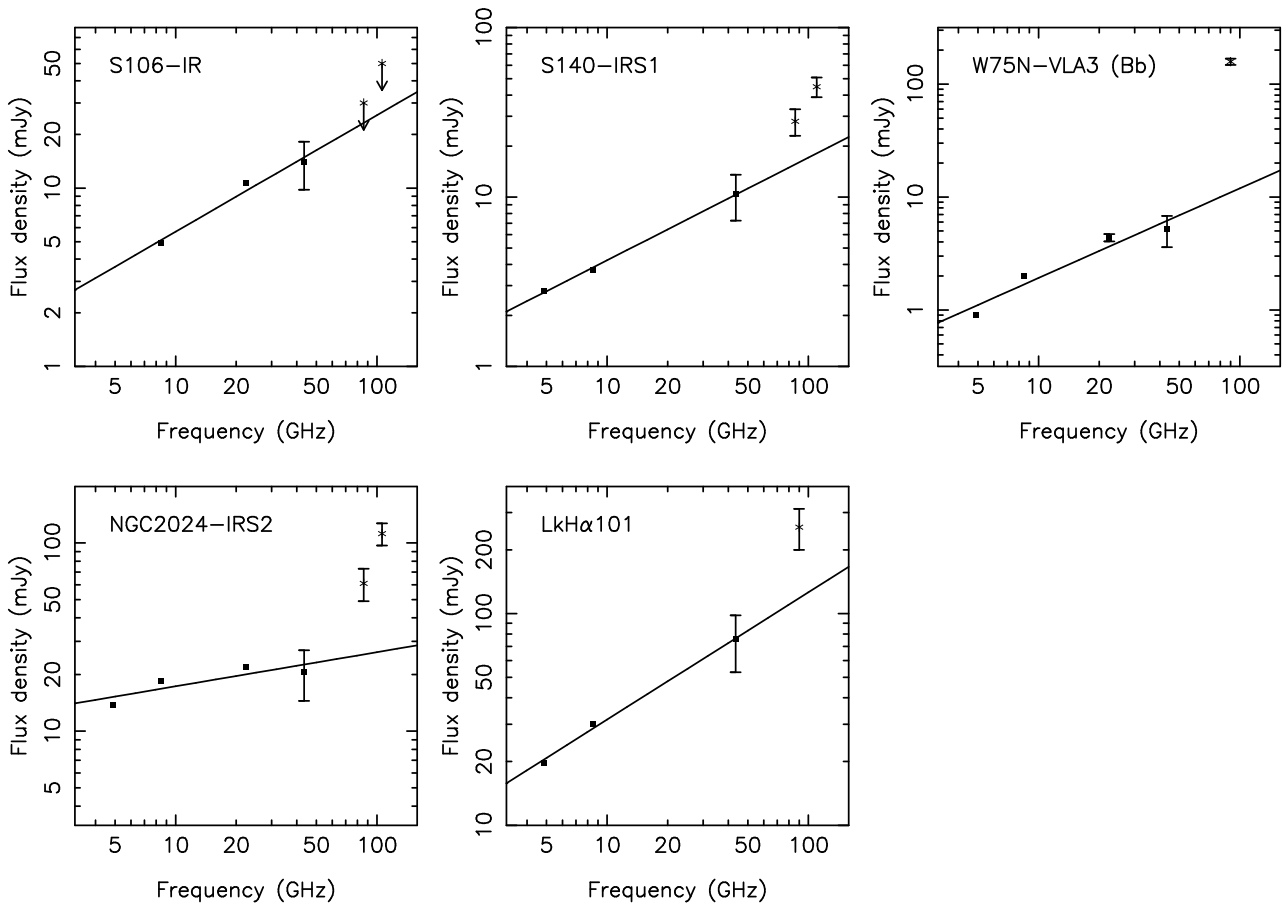}
\caption{ Radio spectral energy distributions for the sources in our
  sample (filled squares). Error bars are drawn where they are greater
  than the symbol size. The 22.5-GHz data points for \lkha\ and
  S140-IRS1 have been omitted as they appear to be significant
  underestimates of the flux density at this frequency.  Asterisks
  mark 3-mm measurements, taken with much larger beams than the
  radio. For NGC\,2024-IRS2 and S\,106-IR (upper limits) these are our
  Plateau de Bure data at 86 and 106 GHz. For S140-IRS1 these are OVRO
  measurements at 86 and 110 GHz (Gibb, Hoare \& Shepherd, in
  preparation). Points for W75N and \lkha\ are taken from Shepherd
  (2001) and Altenhoff, Thum \& Wendker (1994) respectively. }
\label{sed}
\end{figure*}

We derive a spectral index of 0.79$\pm$0.15 for VLA3, although the fit
is rather poor: from Fig. \ref{sed} it is clear that the observed
fluxes do not lie on a straight line. However, our spectral index is
consistent with an ionized wind and the value of 0.5 derived by
Shepherd et al.\ (2004) at the 2-$\sigma$ level. The spectral index
derived by Torrelles et al.\ (1997) is higher at +1.5, although their
estimate comes from non-simultaneous observations. The brightness
temperature of VLA3 increases with frequency, although it is much
lower than the typical temperature of ionized gas and the stellar
effective temperature.

Our spectral index for W75N-Bc is in excellent agreement with that
derived by Shepherd et al.\ (2004). These authors point out that it is
likely that Bc has significant extended flux, which we are filtering
out in our high-resolution observations. The biggest difference lies
with the spectral index for VLA1, for which we calculate a large
negative value, compared with previous positive estimates of 0.2 to
0.7. This is undoubtedly a consequence of our resolving out of
emission at 8.5 GHz compared with the lower-resolution observations of
Hunter et al.\ (1994).

\subsection{NGC2024-IRS2}

\begin{figure*}
\centering
\includegraphics[width=16.5cm]{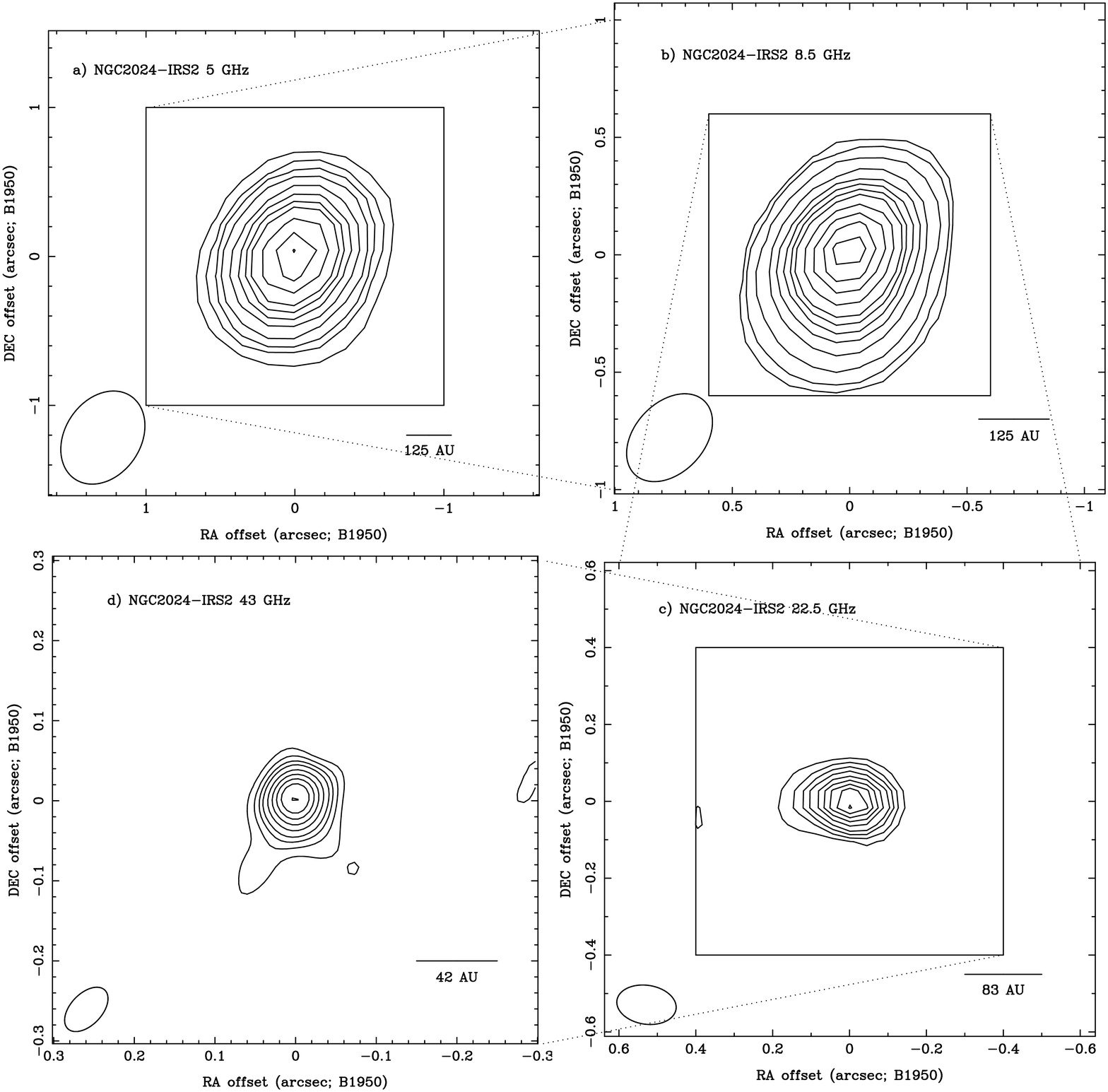}
\caption{Images of NGC\,2024-IRS2 at each frequency, clockwise
  starting from top left: a) 5 GHz, b) 8.5 GHz, c) 22.5 GHz and d) 43
  GHz. Contours are at a) $-$3, 3, 5, 7, 10, 15, 20, 25, 30, 40, 50,
  60, 70$\times$0.20 mJy\,beam$^{-1}$; b) $-$3, 3, 5, 10, 20, 30, 40,
  50, 60, 80, 100, 120, 140, 160$\times$0.091 mJy\,beam$^{-1}$; c)
  $-$3, 3, 5, 7, 9, 12, 15, 18, 21, 25, 30, 35, 40$\times$0.47
  mJy\,beam$^{-1}$; and d) $-$3, 3, 5, 7, 9, 12, 15, 18, 21, 25, 30,
  35, 40$\times$0.40 mJy\,beam$^{-1}$ respectively. The large box in
  a), b) and c) represents the size of the next image. The open
  ellipse in the bottom left corner represents the beam, and a scale
  bar is shown as a horizontal line assuming the distance given in
  Table \ref{sources}. The NE extension on the 22.5-GHz image is a
  residual phase error.}
\label{irs2}
\end{figure*}

NGC\,2024-IRS2 is a bright infrared source in the NGC\,2024 nebula,
which is excited by the nearby source IRS2b (Bik et al.\ 2003).  Deep
VLA observations at 3.6 cm by Rodr\'{\i}guez, Gomez \& Reipurth (2003,
hereafter RGR03) detected both IRS2 and IRS2b, as well as a close
companion to IRS2b. Alvarez et al.\ (2004) detect IRS2 as a point
source with no evidence for secondary companions or extended emission.
IRS2 underwent an outburst between 1989 and 1991 whereby its
near-infrared output increased by a factor of 2.5 (Nisini et al.\
1994); IRS2 is also variable at radio wavelengths (Lenorzer et al.\
2004).

\begin{figure*}
\centering
\includegraphics[width=16.5cm]{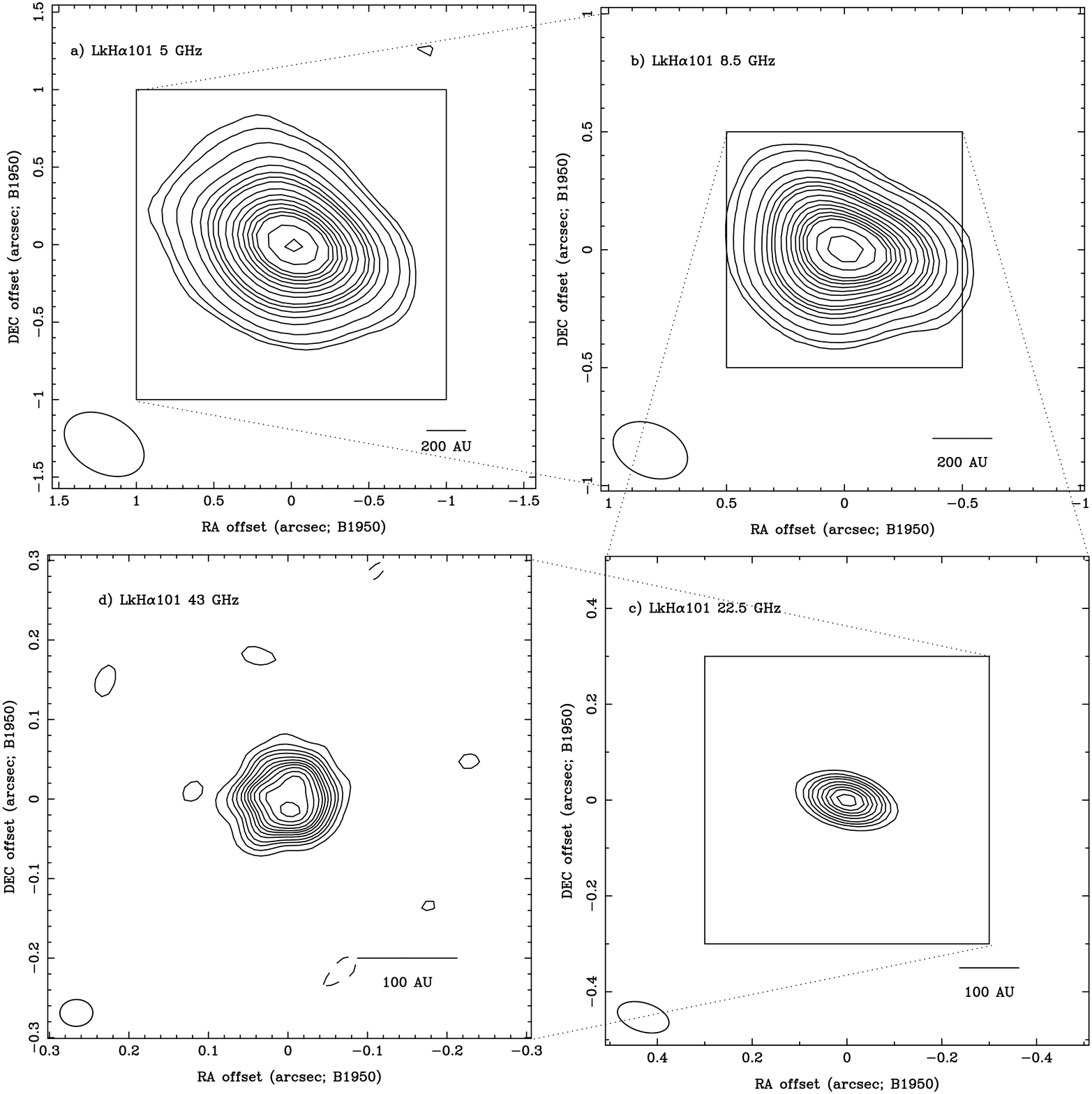}
\caption{Images of \lkha\ at each frequency, clockwise starting from
  top left: a) 5 GHz, b) 8.5 GHz, c) 22.5 GHz and d) 43 GHz. Contours
  are at a) $-$3, 3, 5, 10, 20, 30, 40, 50, 60, 80, 100, 120, 140,
  160, 180, 200, 250, 300$\times$0.043 mJy\,beam$^{-1}$ b) $-$3, 3, 5,
  10, 20, 30, 40, 50, 60, 80, 100, 120, 140, 160, 180, 200, 250, 300,
  350, 400, 450$\times$0.055 mJy\,beam$^{-1}$ c) $-$3, 3, 5, 10, 15,
  20, 25, 30, 35, 40, 45, 50, 55, 60$\times$0.30 mJy\,beam$^{-1}$ and
  d) $-$3, 3, 5, 7, 9, 12, 15, 18, 21, 25, 30, 35$\times$0.31
  mJy\,beam$^{-1}$ respectively. The large box in a), b) and c)
  represents the size of the next image. The open ellipse in the
  bottom left corner represents the beam, and a scale bar is shown as
  a horizontal line assuming the distance given in Table
  \ref{sources}. }
\label{lkha101}
\end{figure*}

IRS2 was imaged at all four wavelengths in our survey, and the results
are shown in Fig. \ref{irs2}. IRS2 is unresolved at all wavelengths,
with the possible exception of 43 GHz where the emission appears
circularly-symmetric despite the elongated beam. The deconvolved
source size indicates an elongated source along an axis perpendicular
to the beam. Since IRS2 is our closest source our resolution at 43 GHz
is probing ionized gas on scales of 17--20 AU from the central
star. The flux we measure for IRS2 at 8.5 GHz (18.5 mJy) is in good
agreement with the recent measurement of RGR03, which they note is
approximately a factor of 2 higher than previous observations (Kurtz,
Churchwell \& Wood 1994). RGR03 also show that IRS2 is slightly
extended to the east at 8.5 GHz, with perhaps a second source 0.37
arcsec ESE of IRS2 which we do not see in any of our images.

The radio spectrum for IRS2 rises only slowly across the range of
frequencies observed here with a fitted spectral index of only
+0.18$\pm$0.08. The spectrum appears to show some flattening above 8.5
GHz, giving rise to a spectrum closer that expected for an \HII\
region. The value of the angular size power-law index, $\zeta$, is
lower than that expected for a uniform wind. The brightness
temperature of IRS2 increases slightly with frequency, decreasing
again at 43 GHz, but the variation is small (see Table \ref{fits}).

We do not detect either of the sources associated with IRS2b (VLA\,15
and 16 in the notation of RGR03). For VLA\,15 this is a sensitivity
issue since its flux density at 8.5 GHz is only 0.24 mJy (RGR03) and
would only be detected at the 3-$\sigma$ level in our map. The
non-detection of VLA\,16 suggests that it may also be time-variable.

We note that our 22-GHz flux is lower than previous measurements.
Lenorzer et al.\ (2004) suggest that IRS2 underwent a period of radio
brightening in 1994-1995 relaxing to lower levels around 2002, a
period for which they have no other data. Our data for 1996 November 1
show that NGC2024-IRS2 must have decreased in brightness rather more
rapidly, reaching pre-outburst fluxes (at least at high frequencies
where the gas will become optically thin first) within a period of 15
months. However, we have made the only 43-GHz measurement to date
which makes it difficult to assess how much the flux at that frequency
has changed. Lenorzer et al.\ (2004) model the radio spectrum pre- and
post-outburst as a recombining stellar wind with modest differences
(of order a factor of 2) in the outer radius and electron density at
the stellar surface.

Bik et al.\ (2003) suggested that IRS2b is the exciting source of the
NGC\,2024 \HII\ region. Comparing our results with RGR03 shows that
the 8.5-GHz flux of IRS2b (VLA\,16 in their terminology) increased by
a factor of at least 3 over the time period 1996 November to 2002
March. Kurtz et al.\ (1994) also did not detect it at 8.5 and 15 GHz
to a 3$\sigma$ level of 0.33 and 2.2 mJy respectively. Thus it may be
that IRS2b is currently undergoing an outburst, and observations are
clearly required to explore this further.

\begin{table*}
\centering
\caption{Fluxes of radio sources detected. Peak ($F_{\rm pk}$) and
  total ($F_{\rm tot}$) flux densities are given in mJy\,beam$^{-1}$
  and mJy respectively. Brightness temperatures ($T_{\rm b}$) are
  derived from the peak flux and are quoted in K. Uncertainties in the
  brightness temperatures are derived from those for the corresponding
  flux density. The tenth column lists the derived spectral
  index. Upper limits are given as 3-$\sigma$ where $\sigma$ is taken
  from Table \ref{obsparms}. The Notes column indicates if only 2
  points have been used to derive the spectral index. }
\begin{tabular}{lcccccc}
Source & $\nu$ & $S_{\rm pk}$ & $S_{\rm tot}$ & $T_{\rm b}$ & $\alpha$ & Notes \\
       & GHz & mJy\,beam$^{-1}$ & mJy & K & & \\
\hline
S\,106-IR         & 8.5  &     4.5$\pm$0.1 &    4.8$\pm$0.3  & 2060$\pm$45    & 0.65$\pm$0.11 & VLA only \\
                  & 22.5 &     7.9$\pm$0.2 &   10.7$\pm$0.4  & 2500$\pm$60    &               &   \\
                  & 43   &     6.7$\pm$0.4 &   14.0$\pm$0.4  & 3530$\pm$210   &               &   \\
                  & 23   &     3.6$\pm$1.0 &   12.1$\pm$2.0  & 12500$\pm$4000 &               & MERLIN \\
\hline
S\,140-IRS1       & 5    &     1.9$\pm$0.1 &    2.8$\pm$0.3  &  580$\pm$30    & 0.61$\pm$0.03 &   \\ 
                  & 8.5  &     1.9$\pm$0.1 &    3.7$\pm$0.2  &  570$\pm$30    &               &   \\
                  & 22.5 &     3.4$\pm$0.2 &    4.2$\pm$0.3  &  970$\pm$60    &               &   \\
                  & 43   &     3.7$\pm$0.4 &   10.4$\pm$0.5  & 1850$\pm$200   &               &   \\
\hline
S\,140-IRS2N      & 5    &     1.5$\pm$0.1 &    1.7$\pm$0.1  &  460$\pm$30    & 0.77          & 2 \\ 
                  & 8.5  &     2.4$\pm$0.1 &    2.6$\pm$0.3  &  720$\pm$30    &               &   \\
                  & 22.5 & $<$0.78         & $<$0.78         &  --            &               &   \\
                  & 43   & $<$1.08         & $<$1.08         &  --            &               &   \\
\hline
W75N-VLA1         & 5    &     1.4$\pm$0.1 &    1.7$\pm$0.1  &  450$\pm$30    & $-$1.60       & 2 \\ 
                  & 8.5  &     0.7$\pm$0.1 &    0.7$\pm$0.1  &  220$\pm$30    &               &   \\
                  & 22.5 & $<$0.96         & $<$0.96         &  --            &               &   \\
                  & 43   & $<$0.90         & $<$0.90         &  --            &               &   \\
\hline
W75N-VLA3         & 5    &     0.7$\pm$0.1 &    0.9$\pm$0.1  &  220$\pm$30    & 0.79$\pm$0.15 &   \\ 
                  & 8.5  &     1.6$\pm$0.1 &    2.0$\pm$0.1  &  490$\pm$30    &               &   \\
                  & 22.5 &     3.5$\pm$0.4 &    3.7$\pm$0.5  & 1040$\pm$120   &               &   \\
                  & 43   &     3.1$\pm$0.3 &    5.2$\pm$0.3  & 1560$\pm$150   &               &   \\
\hline
W75N-Bc           & 5    &     0.5$\pm$0.1 &    0.6$\pm$0.1  &  160$\pm$30    & $-$0.34       & 2 \\ 
                  & 8.5  &     0.4$\pm$0.1 &    0.5$\pm$0.1  &  120$\pm$30    &               &   \\
                  & 22.5 & $<$0.96         & $<$0.96         &  --            &               &   \\
                  & 43   & $<$0.90         & $<$0.90         &  --            &               &   \\
\hline
NGC\,2024-IRS2    & 5    &    12.1$\pm$0.2 &   13.8$\pm$0.1  & 1840$\pm$30    & 0.18$\pm$0.08 &   \\
                  & 8.5  &    13.8$\pm$0.1 &   18.4$\pm$0.2  & 1850$\pm$15    &               &   \\
                  & 22.5 &    17.0$\pm$0.4 &   22.0$\pm$0.5  & 2620$\pm$60    &               &   \\
                  & 43   &    10.0$\pm$0.4 &   20.7$\pm$0.5  & 2500$\pm$100   &               &   \\
\hline
Lk\,H$\alpha$101  & 5    &    13.4$\pm$0.1 &   19.7$\pm$0.5  & 3450$\pm$25    & 0.51$\pm$0.06 &   \\
                  & 8.5  &    18.5$\pm$0.1 &   30.0$\pm$0.6  & 4300$\pm$25    &               &   \\
                  & 22.5 &    16.0$\pm$0.3 &   20.0$\pm$2.0  & 5800$\pm$109   &               &   \\
                  & 43   &    13.1$\pm$0.3 &   62.4$\pm$2.5  & 6110$\pm$140   &               &   \\
\hline
\end{tabular}
\label{fits}
\end{table*}

\subsection{LkH\boldmath $\alpha$101}

\lkha\ is a Herbig Be star at a distance of 700\,pc (Herbig et al.\
2004), associated with an optical reflection nebula (NGC\,1579) and a
compact \HII\ region (S222). Previous VLA observations by Becker \&
White (1988) revealed a compact stellar-wind source within a more
extended shell-like \HII\ region. MERLIN observations by Hoare et
al.\, (1994) and Hoare \& Garrington (1995) at 5 and 1.6 GHz resolved
structure in the stellar wind on scales of $\sim$50 AU, while Bieging,
Cohen \& Schwartz (1984) also noted departures from circular symmetry
in their 5 and 15 GHz VLA images. Recently Alvarez et al.\ (2004) made
near-infrared speckle observations of \lkha\ and detected a secondary
component 0.17 arcsec to the east of the main star. Even higher
resolution infrared observations by Tuthill et al.\ (2002) reveal both
the secondary source and show that \lkha\ has a shell-like appearance
in the 1--3 $\mu$m range, which they modelled as an almost face-on
disc with a central hole.

\lkha\ was imaged at all four wavelengths in our survey, and the
results are shown in Fig. \ref{lkha101}. Since all our observations
used only the VLA A configuration we are not sensitive to emission
from the ionized gas on scales greater than 10 arcsec. In practice our
images contain no emission on scales much larger than the beam itself:
we only detect emission from the central ionized wind rather than the
more extended \HII\ region seen by Becker \& White (1988).  \lkha\ is
resolved at all wavelengths. The 5 and 8.5 GHz images (Figs
\ref{lkha101}a and b) show some asymmetry in the lower contours, and
our 8.5 GHz image bears a strong resemblance to the 15 GHz image of
Bieging et al.\ (1984). However we do not detect their SE extension in
any of our images. Our 5-GHz data are in reasonable agreement with
those of Becker \& White (1988); the total flux in our image is only
about two-thirds of their measurement (although our peak flux is in
good agreement with the measurement of Bieging et al.\ 1984), probably
due to our insensitivity to emission on larger scales.

The new result here comes from our 43 GHz data. The emission retains a
high degree of circular symmetry, even on scales of less than 100
AU. We do not detect the secondary companion to the east.  The
plateau-like appearance in Fig. \ref{lkha101}d is genuine and not an
artefact of the choice of contour levels. Such an appearance is
consistent with an equatorial wind viewed face on and the near face-on
view proposed by Tuthill et al.\ (2002).  We measure a total flux of
62 mJy at this frequency, somewhat less than expected based on
extrapolating the observations of Becker \& White (1988), although we
are probably only detecting the inner regions of the wind. However, we
calculate a peak brightness temperature of $\sim$6000\,K from our 43
GHz measurements, lower than expected if we were detecting emission
from the innermost regions of the wind.

The spectral index for \lkha\ is 0.51$\pm$0.06 (excluding our 22.5-GHz
data), is in agreement with the value expected for an ionized wind at
the 2-$\sigma$ level. If we apply a taper to the 43-GHz data (reducing
the resolution to 0.2 arcsec) then we recover a total flux of
$\sim$75-80 mJy, which then yields a spectral index of
0.61$\pm$0.04. Examining Fig. \ref{sed} shows that the 43 GHz point
lies below a line extrapolated from the 5 and 8.5-GHz points. However,
the A-array VLA data of Becker \& White (1988) yield a 22.5-GHz flux
density of order 60--80 mJy, which implies a 43-GHz flux density of
more than 100 mJy assuming the above spectral index. The most likely
explanation for our apparently anomalously low 22.5-GHz measurement is
our poorer sensitivity. The brightness temperatures increase with
frequency from 3450 to 6110\,K, and are the highest of all our sources
at each frequency. Despite that fact, they are still consistently
below 10\,000\,K and the stellar effective temperature. This may be
explained if the emission becomes optically thick in a relatively cool
region of the wind, or the filling factor is less than 100 per cent.

The MERLIN observations of Hoare et al.\ (1994) show that the wind is
resolved on scales of a few tens of AU at 5 GHz. The more circular
appearance that we obtain at 43 GHz compared with the MERLIN 5-GHz
image can be attributed to the emission being less well-resolved at 43
GHz. The angular size we derive at 43 GHz (74 milliarcsec) is in good
agreement with that predicted from the MERLIN data, assuming that the
size falls of as $\nu^{-0.6}$ (Wright \& Barlow 1975). Therefore, the
ionized gas towards \lkha\ appears to be well-characterized by a
spherically-symmetric, constant velocity wind, a conclusion reinforced
by the results of the modelling described in Section \ref{modresults}.

There are a number of possible reasons for lower-than-expected flux
densities at the highest frequencies. One explanation which applies to
the VLA (or any interferometer) in its most extended configuration
operating at the highest frequencies is decorrelation due to phase
noise, which leads to a reduction in the visibility amplitude on the
longest baselines. However, plots of the calibrator visibility
amplitudes showed that the fast-switching technique had successfully
stopped the atmosphere and decorrelation was negligible. Another
reason may be our absolute calibration at 43 GHz, since we assumed,
rather than derived, a flux density for our primary calibrator.  The
most likely explanation is that we have resolved the wind at 43 GHz
and the more extended emission is too weak to be detected by our
observations. However poor sampling at short baselines can also result
in a significant underestimate of the total flux (Wilner \& Welch
1994).


\begin{table*}
\centering
\caption{Observed and deconvolved sizes of radio sources at each
  detected wavelength. Columns 3--5 list the parameters derived from
  the fits while columns 6--8 list the parameters for a gaussian
  deconvolution. The angular size-frequency power law index, $\zeta$,
  is listed in column 9. Two values are listed for each source: the
  first value is from a fit to the convolved (fitted) sizes, the
  second value is from a fit to the deconvolved sizes. The Notes
  column indicates if only 2 points have been used in the fit for
  $\zeta$. In some cases the error in the deconvolved solution is as
  large as the minor axis.}
\begin{tabular}{lccccccccccc}
Source & $\nu$ & $\theta_{\rm a}$ & $\theta_{\rm b}$ & PA  & $\theta_{\rm a}'$ & $\theta_{\rm b}'$ & PA  & $\zeta$ & Notes \\
       & GHz   &  arcsec          &  arcsec          & deg &   arcsec          &  arcsec           & deg &         & \\
\hline
S\,106-IR         &  8.5 &  0.25$\pm$0.02  &  0.18$\pm$0.02  & 153$\pm$4  &  0.04$\pm$0.04  &  0.02$\pm$0.02  & 150$\pm$30 &  $-$0.95$\pm$0.14  &  \\
                  & 22.5 & 0.105$\pm$0.001 & 0.079$\pm$0.006 &  71$\pm$1  & 0.025$\pm$0.025 &  0.01$\pm$0.01  & 143$\pm$0  &  $-$0.45$\pm$0.24  &  \\
                  & 43   & 0.049$\pm$0.009 & 0.034$\pm$0.003 & 107$\pm$3  & 0.031$\pm$0.010 & 0.010$\pm$0.009 & 129$\pm$28 &  &  \\
\hline
S\,140-IRS1       &  5   &  0.60$\pm$0.01  &  0.42$\pm$0.01  & 169$\pm$1  &  0.35$\pm$0.02  &  0.11$\pm$0.05  &  34$\pm$2  &  $-$0.93$\pm$0.05  &  \\
                  &  8.5 &  0.40$\pm$0.01  &  0.27$\pm$0.01  &  16$\pm$1  &  0.32$\pm$0.01  &  0.03$\pm$0.03  &  37$\pm$1  &  $-$0.69$\pm$0.27  &  \\
                  & 22.5 & 0.122$\pm$0.009 & 0.077$\pm$0.006 & 169$\pm$6  & 0.033$\pm$0.014 & 0.025$\pm$0.025 &  27$\pm$40 &  &  \\
                  & 43   & 0.081$\pm$0.007 & 0.048$\pm$0.004 &  70$\pm$6  & 0.074$\pm$0.006 & 0.026$\pm$0.007 &  66$\pm$4  &  &  \\
\hline
S\,140-IRS2N      &  5   &  0.60$\pm$0.02  &  0.03$\pm$0.01  & 157$\pm$2  &  0.22$\pm$0.06  &  0.06$\pm$0.06  & 171$\pm$12 &  $-$0.95$\pm$0.14  & 2 \\
                  &  8.5 &  0.32$\pm$0.01  &  0.19$\pm$0.01  & 156$\pm$2  &  0.06$\pm$0.02  &  0.02$\pm$0.02  &  80$\pm$40 &  $-$0.45$\pm$0.04  & 2 \\
\hline
W75N-VLA1         &  5   &  0.50$\pm$0.02  &  0.37$\pm$0.01  & 145$\pm$5  &  0.17$\pm$0.03  &  0.08$\pm$0.03  &  53$\pm$41 &  $-$1.11$\pm$0.02  & 2 \\
                  &  8.5 &  0.30$\pm$0.01  &  0.31$\pm$0.01  & 141$\pm$5  &  0.12$\pm$0.03  &  0.05$\pm$0.05  & 110$\pm$26 &  $-$0.97$\pm$0.21  & 2 \\
\hline
W75N-VLA3         &  5   &  0.50$\pm$0.10  &  0.36$\pm$0.03  & 145$\pm$2  &  0.33$\pm$0.01  &  0.10$\pm$0.10  & 120$\pm$20 &  $-$1.05$\pm$0.08  &  \\
                  &  8.5 &  0.32$\pm$0.01  &  0.20$\pm$0.01  & 144$\pm$2  &  0.13$\pm$0.01  &  0.06$\pm$0.01  & 141$\pm$2  &  $-$0.75$\pm$0.30  &  \\
                  & 22.5 & 0.141$\pm$0.012 & 0.075$\pm$0.006 &   2$\pm$5  & 0.094$\pm$0.015 & 0.002$\pm$0.002 &   4$\pm$4  &  &  \\
                  & 43   & 0.048$\pm$0.004 & 0.043$\pm$0.004 & 131$\pm$33 & 0.031$\pm$0.007 & 0.022$\pm$0.008 &   0$\pm$20 &  &  \\
\hline
W75n-Bc           &  5   &  0.67$\pm$0.07  &  0.41$\pm$0.04  & 138$\pm$8  &  0.43$\pm$0.08  &  0.24$\pm$0.01  & 131$\pm$18 &  $-$1.21$\pm$0.41  & 2 \\
                  &  8.5 &  0.30$\pm$0.03  &  0.27$\pm$0.02  & 116$\pm$41 &  0.20$\pm$0.20  &  0.03$\pm$0.03  &  67$\pm$13 &  $-$2.00$\pm$0.35  & 2 \\
\hline
NGC\,2024-IRS2    &  5   &  0.69$\pm$0.02  &  0.54$\pm$0.02  & 148$\pm$2  &  0.22$\pm$0.03  &  0.10$\pm$0.10  &  60$\pm$40 &  $-$1.00$\pm$0.07  &  \\
                  &  8.5 &  0.44$\pm$0.02  &  0.33$\pm$0.03  & 143$\pm$7  &  0.23$\pm$0.23  &  0.11$\pm$0.11  &  17$\pm$17 &  $-$0.86$\pm$0.89  &  \\
                  & 22.5 & 0.165$\pm$0.010 & 0.110$\pm$0.010 &  82$\pm$1  & 0.081$\pm$0.081 & 0.069$\pm$0.069 &  92$\pm$92 &  &  \\
                  & 43   & 0.072$\pm$0.007 & 0.055$\pm$0.015 & 148$\pm$12 & 0.045$\pm$0.020 & 0.019$\pm$0.019 &  17$\pm$15 &  &  \\
\hline
Lk\,H$\alpha$101  &  5   & 0.640$\pm$0.002 & 0.470$\pm$0.001 &  64$\pm$14 & 0.330$\pm$0.003 & 0.290$\pm$0.002 &  84$\pm$2  &  $-$0.86$\pm$0.01  &  \\
                  &  8.5 & 0.420$\pm$0.002 & 0.290$\pm$0.001 &  74$\pm$1  & 0.260$\pm$0.002 & 0.180$\pm$0.001 &  86$\pm$1  &  $-$0.68$\pm$0.02  &  \\
                  & 22.5 & 0.121$\pm$0.002 & 0.065$\pm$0.001 &  76$\pm$1  & 0.047$\pm$0.001 & 0.024$\pm$0.001 &  86$\pm$6  &  &  \\
                  & 43   & 0.085$\pm$0.003 & 0.075$\pm$0.002 & 115$\pm$11 & 0.074$\pm$0.005 & 0.066$\pm$0.002 & 122$\pm$13 &  &  \\
\hline
\end{tabular}
\label{sizes}
\end{table*}

\section{Modelling the radio emission as a uniform spherical wind}

\subsection{Model description}
\label{modelling}

\begin{figure}
\centering
\includegraphics[width=7.5cm]{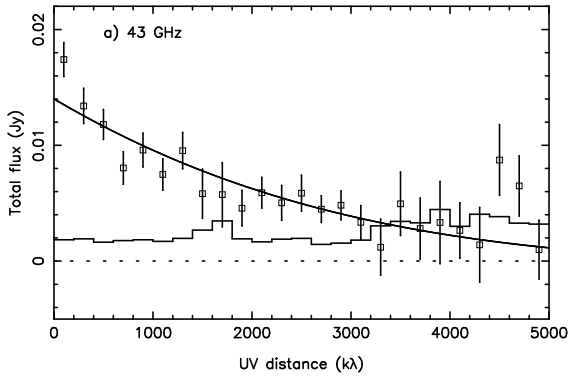}
\caption{S106-IR visibility amplitude for 43 GHz data. The amplitude
is in Jy and the baseline is in kilo-wavelengths. The error bars are
$\pm$1-$\sigma$. The best-fitting model from Table \ref{modelfits} is
also shown as a solid curve. The histogram in this and subsequent
figures denotes the expected flux (due to noise alone) in the presence
of zero signal. \label{s106uv}}
\end{figure}

In order to explore the nature of the ionized emission, we have
followed the analysis of Becker \& White (1982, hereafter BW82) and
modelled the emission from a uniform temperature spherical stellar
wind. As these authors argue, for stellar wind sources with extended
faint emission it is better to analyze interferometer data in the
spatial-frequency domain rather than the image domain. In their model,
the only free parameters are the total flux (i.e.\ flux at zero
baseline length, $S_{\rm T}$) and the wind temperature ($T_{\rm
w}$). The slope of the visibility function constrains the wind
temperature, while the extrapolated zero-spacing flux constrains the
total flux or equivalently, $\dot{M}/v_\infty$, the mass-loss rate
divided by the wind terminal velocity.

Table \ref{modelfits} lists the best-fitting parameters for each
source. Table \ref{modelmdot} lists a number of properties derived
from the models, including $\dot{M}/v_\infty$ which we convert to a
mass-loss rate assuming a wind terminal velocity where known
(typically in the range 200--400\,\kms: see, e.g.\ Drew et al.\ 1993;
Chandler, Carlstrom \& Scoville 1995; Bunn et al.\ 1995; Lenorzer et
al.\ 2004). These mass-loss rates are not very sensitive to the
adopted wind temperature but obviously scale directly with the wind
velocity.

\begin{figure}
\centering
\includegraphics[width=7.5cm]{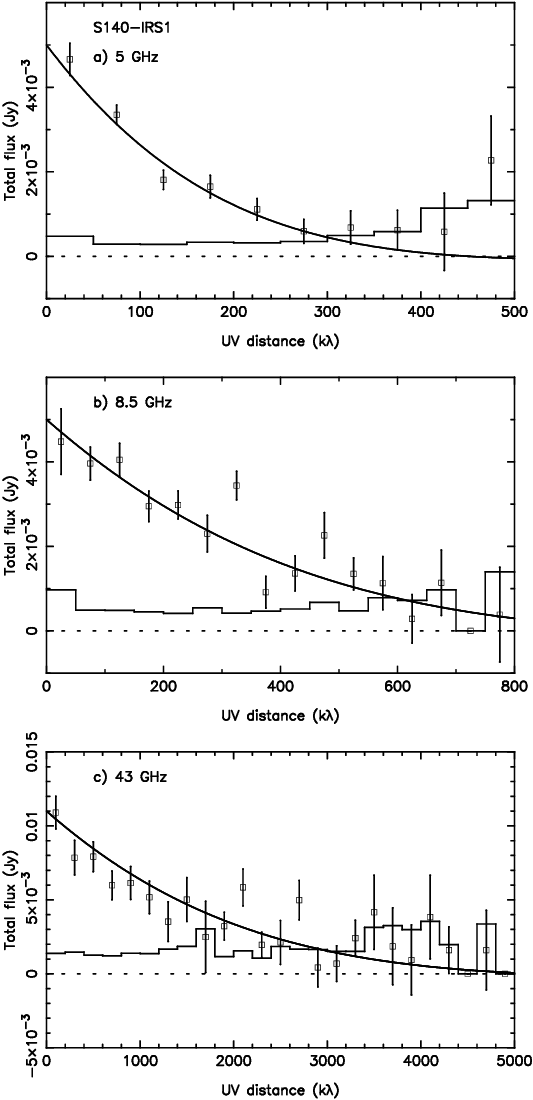}
\caption{S140-IRS1 visibility amplitude for 5, 8.5 and 43 GHz. The
amplitude is in Jy and the baseline is in kilo-wavelengths. The error
bars are $\pm$1-$\sigma$. The best-fitting models from Table
\ref{modelfits} is also shown as a solid curve. \label{s140uv}}
\end{figure}

\begin{figure*}
\centering
\includegraphics[width=15cm]{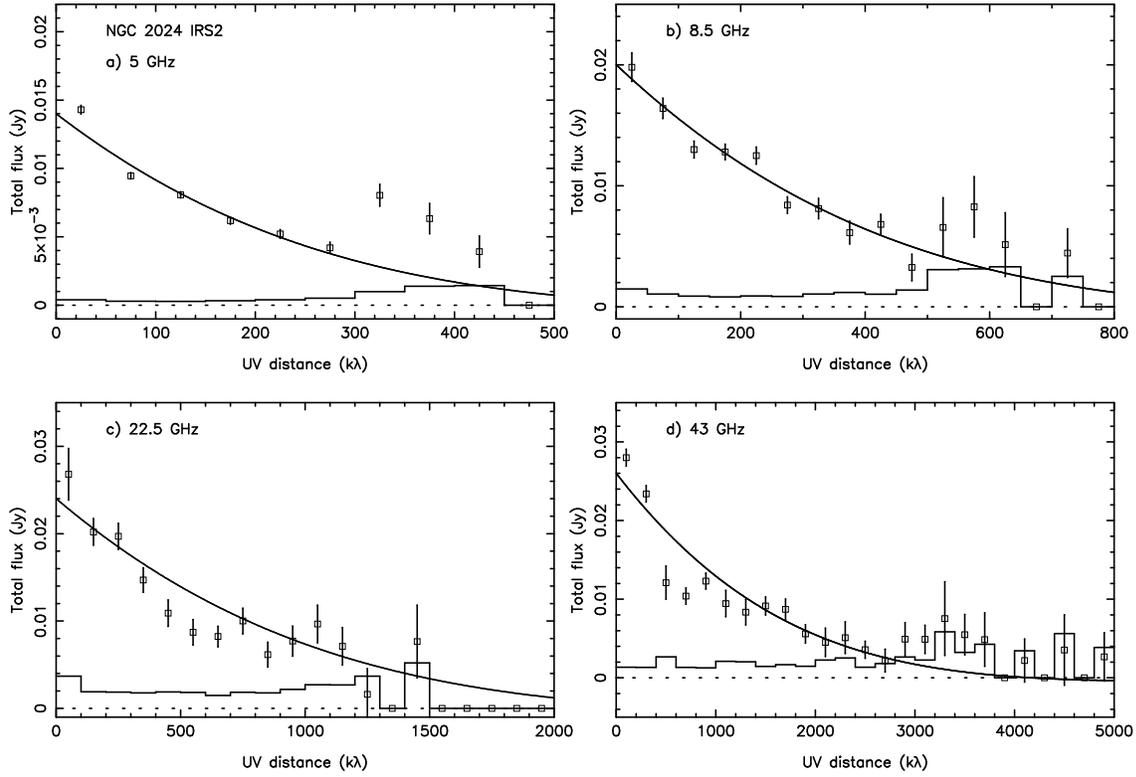}
\caption{NGC2024\,IRS2 visibility amplitude for each observed
wavelength. The amplitude is in Jy and the baseline is in
kilo-wavelengths. The error bars are $\pm$1-$\sigma$. The best-fitting
model from Table \ref{modelfits} is also shown as a solid
curve.\label{n2024uv}}
\end{figure*}

\begin{figure*}
\centering
\includegraphics[width=15cm]{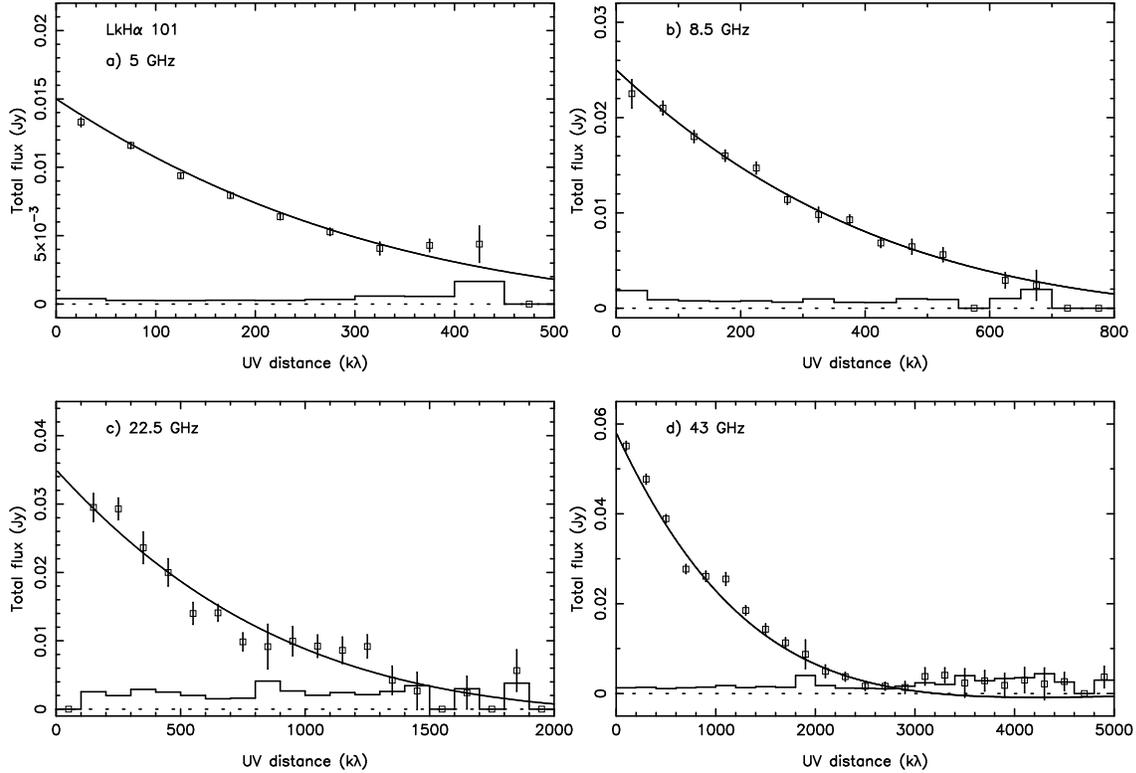}
\caption{\lkha\ visibility amplitude for each observed wavelength. The
amplitude is in Jy and the baseline is in kilo-wavelengths. The error
bars are $\pm$1-$\sigma$. The best-fitting model from Table
\ref{modelfits} is also shown as a solid curve. 
\label{lkha101uv}}
\end{figure*}

Figs  \ref{s106uv} to \ref{lkha101uv}  show plots of the
azimuthally-averaged visibility amplitude as a function of baseline
length for each source, with the best fitting spherical wind models
overlaid. The error bars represent the 1-$\sigma$ noise level and the
histogram on each plot represents the expected flux for zero signal
based on the noise level. When determining the goodness of fit, data
points less than 1$\sigma$ different from the zero-signal flux were
excluded. 

Good fits were obtained if the value of $\chi^2$ passed through a
well-defined minimum. In general the total fluxes were reasonably
well-constrained, often exhibiting well-defined minima in the $\chi^2$
distribution. Reasonable fits could typically be obtained with a wide
range of temperatures (typically 2000 to 5000\,K).

The best-fitting flux at higher temperatures tended to be lower than
that at lower temperatures. This is easy to understand. For a given
flux density, at high temperatures the model visibility function falls
off more slowly than at low temperatures. Thus in order to yield a
`good' minimum-$\chi^2$ fit, the total flux must decrease.

\begin{table*}
\centering
\caption{Best-fitting spherically-symmetric, isothermal wind model
  parameters for each frequency. The total flux is in mJy and the wind
  temperature is in K. The final column gives the spectral index
  ($\alpha_{\rm mod}$) corresponding to the best-fitting fluxes at
  each frequency.
  \label{modelfits}}
\begin{tabular}{lccccccccc}
Source &  \multicolumn{2}{c}{5 GHz}  &
\multicolumn{2}{c}{8.5 GHz}  &
\multicolumn{2}{c}{22.5 GHz}  &
\multicolumn{2}{c}{43 GHz}      & $\alpha_{\rm mod}$ \\
& $S_{\rm T}$ & $T_{\rm w}$ & $S_{\rm T}$ & $T_{\rm w}$ &
       $S_{\rm T}$ & $T_{\rm w}$ & $S_{\rm T}$ & $T_{\rm w}$ & \\
\hline
S106-IR        & --  &   --  & --  & -- & -- & -- & 14.0$\pm$1.0 & 5000$\pm$1000 & --- \\
S140-IRS1      & 5.0$^{+0.5}_{-1.5}$ &  500$^{+1500}_{-500}$ &  5.0$\pm$1.0  & 1000$^{+2000}_{-500}$ & -- & -- & 10.0$\pm$2.0 & 2000$^{+2000}_{-500}$ & 0.35$\pm$0.15 \\
NGC\,2024-IRS2 & 14.0$\pm$2.0  &  3000$\pm$1000 & 20.0$\pm$3.0  & 4000$\pm$1000 & 24.0$\pm$3.0 &  4000$\pm$1000 & 26.0$\pm$3.0 & 3000$\pm$1000 & 0.27$\pm$0.07 \\
\lkha\         & 14.5$\pm$1.5  &  5000$\pm$1000 & 25.0$\pm$2.0 & 5000$\pm$1000 & 35.0$\pm$5.0 &  4500$\pm$1500 & 58.0$\pm$2.0 & 4000$^{+1500}_{-500}$ & 0.59$\pm$0.08 \\
\hline
\end{tabular}
\end{table*}

\begin{table*}
\centering
\caption{Properties derived from the best-fitting wind models listed
  in Table \ref{modelfits} for each frequency. The notation $a(b)$
  denotes $a \times 10^b$.
  \label{modelmdot}}
\begin{tabular}{lcccccc}
Source & $\nu$ & $\dot{M}/v_\infty$ & $q_1$ & $\theta_1$ & $\eta_{\rm
       ff}$ & $\dot{M}$ \\
       & GHz & $M_\odot\,{\rm yr}^{-1} \,({\rm km\,s}^{-1})^{-1}$ & AU
       & arcsec & & $M_\odot\,{\rm yr}^{-1}$ \\
\hline
S106-IR        & 43  & 5.2$\pm$0.4$\times 10^{-9}$       & 9.8$^{+1.5}_{-1.2}$   & 0.016$^{+0.003}_{-0.002}$ & 0.21$^{+0.05}_{-0.06}$ & 1.8$\times 10^{-6}$ \\
\hline
S140-IRS1      & 5   & 1.4$^{+0.2}_{-0.4}\times 10^{-8}$ & 239$^{+11}_{-139}$    & 0.26$^{+0.02}_{-0.15}$    & 0.40$^{+0.06}_{-0.33}$ & 4.2$\times 10^{-6}$ \\
               & 8.5 & 1.0$\pm$0.2$\times 10^{-8}$       & 100$^{+54}_{-48}$     & 0.11$^{+0.06}_{-0.05}$    & 0.21$^{+0.29}_{-0.15}$ & 3.0$\times 10^{-6}$ \\
               & 43  & 8.0$^{+1.4}_{-1.6}\times 10^{-9}$ & 19.7$^{+5.2}_{-7.3}$  & 0.022$^{+0.005}_{-0.008}$ & 0.37$^{+0.19}_{-0.22}$ & 2.4$\times 10^{-6}$ \\
\hline
NGC\,2024-IRS2 & 5   & 8.4$\pm$1.1$\times 10^{-9}$       & 76$^{+24}_{-15}$      & 0.18$^{+0.06}_{-0.04}$    & 0.09$^{+0.07}_{-0.04}$ & 1.7--3.4$\times 10^{-6}$ \\
               & 8.5 & 8.4$^{+1.2}_{-1.1}\times 10^{-9}$ & 46.4$^{+11.1}_{-8.1}$ & 0.11$^{+0.03}_{-0.02}$    & 0.09$^{+0.06}_{-0.03}$ & 1.7--3.4$\times 10^{-6}$ \\
               & 22.5& 6.2$\pm$0.7$\times 10^{-9}$       & 19.2$^{+4.3}_{-3.2}$  & 0.046$^{+0.010}_{-0.008}$ & 0.15$^{+0.07}_{-0.05}$ & 1.2--2.4$\times 10^{-6}$ \\
               & 43  & 5.1$\pm$0.6$\times 10^{-9}$       & 12.1$^{+3.5}_{-2.3}$  & 0.029$^{+0.008}_{-0.006}$ & 0.34$^{+0.21}_{-0.13}$ & 1.0--2.0$\times 10^{-6}$ \\
\hline
\lkha\         & 5   & 2.2$\pm$0.2$\times 10^{-8}$       & 115$^{+20}_{-16}$     & 0.14$^{+0.03}_{-0.02}$    & 0.10$^{+0.05}_{-0.03}$ & 6.6$\times 10^{-6}$ \\
               & 8.5 & 2.6$\pm$0.2$\times 10^{-8}$       & 88.4$^{+12.8}_{-8.3}$ & 0.11$^{+0.02}_{-0.01}$    & 0.17$^{+0.07}_{-0.03}$ & 7.8$\times 10^{-6}$ \\
               & 22.5& 2.2$\pm$0.3$\times 10^{-8}$       & 41.7$^{+12.8}_{-8.3}$ & 0.052$^{+0.016}_{-0.010}$ & 0.30$^{+0.22}_{-0.11}$ & 6.6$\times 10^{-6}$ \\
               & 43  & 2.4$\pm$0.1$\times 10^{-8}$       & 29.8$^{+2.5}_{-4.8}$  & 0.037$^{+0.003}_{-0.006}$ & 0.98$^{+0.17}_{-0.29}$ & 7.2$\times 10^{-6}$ \\
\hline
\end{tabular}
\end{table*}

\subsection{Modelling results for each source}
\label{modresults}

The main result from this modelling is that the fitted wind
temperatures are low, typically 3000--4000\,K, much less than the
effective temperatures for main-sequence stars with the luminosities
given in Table \ref{sources} (22\,000 to 30\,000\,K). They are also
much less than half the effective temperature, which has been shown to
be an appropriate value for the wind temperature in main-sequence O
stars (Drew 1989). However, they are not unprecedently low: Becker \&
White (1988) derived a temperature in the range 6800 to 7800\,K for
\lkha, about 25 per cent higher than the values we derived.

For \lkha\ the solutions at 22.5 and 43 GHz have wind temperatures
which are lower than the observed brightness temperatures at these
frequencies. However, given the absolute calibration uncertainty at
these frequencies (see Section \ref{observations}) this discrepancy is
probably not significant. \lkha\ was quite well-fitted by a uniform,
isothermal spherical wind at all frequencies, as was
NGC\,2024-IRS2. Reasonably good fits were obtained for S106-IR and
S140-IRS1 at 43 GHz, but W75N-VLA3 was not well fitted at any
frequency. We now discuss the results for each source in turn.

\textit{S106-IR:} The 8.5 and 22.5-GHz data could not be fitted with
any temperature. At 43 GHz the fit was quite well-constrained in total
flux of 14 mJy for temperatures in the range 4000 to 6000\,K.

\textit{S140-IRS1:} At 5 GHz the best fitting flux was 5 mJy for a
temperature of 500\,K. Reasonable fits were possible in the range 3.5
to 5.5 mJy and temperatures between 500 and 2000\,K. At 8.5 GHz the
best fit occurred at 1000\,K, also for a total flux of 5 mJy. As with
S106-IR, no fits were possible at 22.5 GHz. At 43 GHz the total flux
was constrained to be $\sim$10\,mJy. The temperature was less well
constrained and reasonable fits in the range 1500 to 4000\,K were
possible.

\textit{NGC2024-IRS2:} At 5 GHz the flux was quite well-constrained at
14 mJy. The corresponding temperatures were in the range 2000 to
4000\,K. The fits at 8.5 GHz were moderately-well constrained with a
best-fitting flux of 20 mJy and temperatures of 3000 to 5000\,K.  Fits
were possible at 22.5 GHz although they were relatively poorly
constrained with in the range 20--27 mJy for temperatures between 3000
and 5000 K. At 43 GHz the best-fitting flux was $\sim$26 mJy for
temperatures between 2000 and 4000\,K.

\textit{\lkha :} The best-constrained fits were obtained for the
brightest source, \lkha. At 5 GHz the best-fit flux was 14.5 mJy with
a temperature around 5000\,K, although temperatures up to 8000\,K gave
reasonable fits. The 8.5-GHz fit showed a similar behaviour with a
flux of 25 mJy and similar temperatures to 5 GHz. The fit at 22.5 GHz
was not as well-constrained in flux (30--40 mJy) but quite well fitted
with a temperature of 4500\,K (plus or minus 1500\,K). At 43 GHz the
flux was not especially well-constrained with best-fit values in the
range 56--60 mJy. The wind temperature was $\sim$4000\,K, although
temperatures up to 5500\,K were almost as good.

\textit{W75N-VLA3:} This source was too faint to be fitted at any
wavelength for any temperature.

\subsection{Lower-than-expected temperatures: clumpy winds?}

As noted above, the model wind temperatures (Table \ref{modelfits})
and the observed brightness temperatures (Table \ref{fits}) are
considerably lower than the values that might be expected for stars of
these luminosities. Unlike Becker \& White (1988), we see no trend for
higher fitted temperatures at higher frequencies, despite the tendency
in the observed brightness temperature estimates (Table
\ref{fits}). The uncertainties in the best-fitting model temperature
do not rule out such a relation, however.

The emission from the model wind of BW82 is dominated by an optically
thick core whose size varies with frequency. If the filling factor for
this optically thick core is less than unity the true brightness
temperature will be underestimated. From equation 5 of BW82 we can
estimate the angular radius at which the wind emission becomes
optically thick for each source. These values were calculated for each
frequency for which we obtained a reasonable model fit, and are listed
in Table \ref{modelmdot}. Comparing these values with the angular
sizes derived in Table \ref{sizes} shows that in many cases our
observations are probing the regime where the wind becomes optically
thick. The corresponding filling factors ($\eta_{\rm ff} =
\theta_1^2/\theta^2$, where $\theta_1$ is the angular diameter of the
$\tau = 1$ surface given in Table \ref{modelfits} and $\theta$ is the
geometric mean of the beam major and minor axes in Table
\ref{obsparms}) are in the range 0.2 to unity at 43 GHz implying that
the observed brightness temperatures at that frequency tend to
underestimate the actual gas temperatures. At lower frequencies
$\eta_{\rm ff}$ tends to be much less than one. For \lkha\ the model
yields wind temperatures which are in reasonable agreement in the
range 6100 to 7500\,K, reinforcing the conclusion that this source is
well-modelled by a spherical isothermal wind. The agreement between
the wind temperature estimates at each frequency is poorer for all the
other sources.
In reality, the stellar wind is unlikely to be smooth as
assumed by the modelling and thus the low filling factors are attained
by the presence of clumps in the flow.

The appearance of S106-IR and S140-IRS1 (Figs \ref{s106} and
\ref{s140} respectively) suggests that the wind is not spherically
symmetric. Clearly such a wind would not completely fill the telescope
beam and it is possible that a non-spherical geometry could mimic a
lower-temperature spherical flow but this requires more detailed 2-D
modelling and is beyond the scope of this paper.

Finally, two points related to interferometers are worth mentioning
here. Decorrelation of the signal on long baselines reduces the
measured flux, which might lead us to underestimate the temperature in
our models. Decorrelation affects high frequency observations the
most, which would show up as lower-than-expected temperatures at high
frequencies. No such effect is seen in our observations, and there is
no evidence for significant decorrelation in our data. The other point
to note is that if we are missing extended flux, its recovery will not
lead to higher-temperature fits: higher temperatures can only be
obtained from the modelling if we are underestimating the flux on the
{\em smallest} scales.

\section{Discussion: Radio emission from massive YSOs}

Together the high frequency observations show that the rising thermal
spectrum continues to 43 GHz in these massive YSOs. There is no
evidence of the need for a dust contribution to the 43 GHz flux. For
the sources with 90--110 GHz measurements there is a significant dust
contribution at these higher frequencies. Even here this can disappear
at high resolution when the flux at these frequencies can also be
dominated by just the ionized wind (Plambeck et al.\ 1995; Gibb et
al.\ 2004).

The origin of the +0.6 spectral index is a consequence of the
contribution of both optically thin and thick emission (Wright \&
Barlow 1975; Panagia \& Felli 1975). Schmid-Burgk (1982) showed that
even non-symmetric winds will have a spectral index similar to the
spherical case provided the electron density falls off as an
inverse-square law. Finite (i.e.\ ionization-bounded) winds will show
higher indices as the emission takes on the appearance of an optically
thick \HII\ region (Simon et al.\ 1983). Jets can have lower or higher
values depending on the properties (Reynolds 1986), though detailed
calculations of collimated winds lean towards values between +0.6 and
+1, depending on the gradient in the electron density (Martin 1996).
Models of photo-evaporating discs by Kessel, Yorke \& Richling (1998)
also predict spectral indices close to that expected for a spherical
ionized wind, irrespective of inclination. Recently, Lugo, Lizano \&
Garay (2006) fitted the centimetre-wave spectrum of NGC\,7538-IRS1 and
MWC349A with a photo-evaporating disc model.

Within the uncertainties in our spectral index calculations, the winds
from four of the five sources are consistent with constant velocity
spherical flows. The spectral index for W75N-VLA3 is slightly larger
and suggests that the wind may be recombining and/or accelerating. The
spectral indices are also not consistent with the winds being
collimated into bipolar jets. However, the values predicted by
Reynolds (1986) are for an unresolved source; since we are resolving
the emission, we are more likely to be dominated by the `core' in his
model which has a spherical-wind geometry.

Since we {\em are} resolving the wind, this does raise the question of
the validity of the calculated spectral indices since the beams at
different frequencies will be sampling different volumes of ionized
gas. Na\"{\i}vely it might be expected that the effect of this would
be to artificially flatten the spectrum, thus implying that the
spectral indices are lower limits. However, the modelling of the
visibility data mitigates the usual problem of matching beams since
the fitted flux is extrapolated to a zero-length baseline. Despite the
large uncertainties, the model spectral indices are all consistent
with ionized winds.

43 GHz observations of massive YSOs are also presented by Menten and
van der Tak (2004) (CRL\,2136), van der Tak and Menten (2005) (W33A,
GL\,2591, NGC\,7538-IRS9), van der Tak et al.\ (2005) (W3-IRS5) and
Reid et al. (2007) (Orion-I). The former find lower brightness
temperatures and their C array fluxes and sizes are larger than their
A array ones. This is ascribed to a centrally-peaked brightness
distribution which is expected for a stellar wind and motivated our
visibility-fitting approach. The sources in W3-IRS5 are mostly
unresolved, but similar in flux and size to those presented here.
They possibly resolve the 43 GHz emission from CRL\,2136 and
NGC\,7538-IRS9 and the derived position angle of the elongation would
align with the bipolar CO flow and therefore be jet-like. A similar
alignment is seen in their C array data on GL\,2591, consistent with
the observation of a jet at 22 GHz by Trinidad et al.\ (2003),
although intriguingly the higher resolution data has a different
orientation and perhaps origin.

The main 43 GHz source observed in W33A is marginally extended in the
C array data and possibly in the higher resolution A array data,
although the derived position angles are very different, rather like
GL\,2591. For W33A there is no well-characterised bipolar CO flow, but
2MASS and Spitzer GLIMPSE images clearly show a large monopolar nebula
extending to the SE, which likely has an outflow origin. The C array
43 GHz emission aligns with this within the uncertainties of the
position angle and hence, may be a jet-like ionized flow.
Furthermore, the interferometric observations of the 106 GHz and 233
GHz continuum emission from W33A-MM1 by van der Tak et al.\ (2000)
appear to be elongated with a position angle of about 50--60
degrees. This is similar to that for the 43-GHz A array result. It is
therefore interesting to speculate whether in this source (and
possibly GL\,2591) that a jet-like wind dominates the flux and
morphology at the lower resolution and a dust disk or torus
perpendicular to the wind is taking over at higher resolution. The SED
at high resolution indicates that this is indeed where the
contributions of wind and dust are comparable.

van der Tak and Menten (2005) chose to interpret their 43 GHz data in
terms of the gravitationally trapped H II region picture of Keto
(2002). However, the sizes and fluxes are similar to those presented
here and consistent with a stellar wind interpretation. The spectral
indices of around unity can also be accommodated in a recombining
wind picture.  W33A and other sources with similar spectral indices
such as GL\,490 (Campbell et al.\ 1986) are also observed to have very
broad (few 100\,\kms) near-IR H I emission lines which again are
consistent with a stellar wind interpretation (Bunn et al.\
1995). Although the modelling of such winds in terms of geometry and
assumptions still has some way to go to match all the observational
data (e.g H\"{o}flich \& Wehrse 1987; Sim et al. 2005), it is clear
that a common origin for the IR line and radio continuum emission is
very likely.

Another motivation for van der Tak and Menten (2005) to consider the
gravitationally trapped H II region picture was to try and account for
the lack of a strong, bright UCHII region powered by Lyman continuum
radiation from these luminous sources. However, a stellar wind can
also help here by reprocessing the hard radiation in the dense regions
close to the star. For the mass-loss rates required to explain the
radio emission it is predicted that the stellar wind would recombine
(H\"{o}flich \& Wehrse 1987) and hence very little Lyman continuum
radiation would emerge into the surroundings. This also explains the
spectral indices somewhat steeper than +0.6 that are often found
(Moran et al.\ 1983). Such models would be a much better fit to the
radio spectra found by van der Tak and Menten than the H II region
models they present. To be effective in suppressing the ionizing
continuum the wind would have to cover all directions. This does not
seem feasible when these sources also power bipolar outflows that are
also likely to originate from close to the star/inner disc. An
alternative explanation is that ongoing accretion keeps the star in a
swollen, low surface gravity, low effective temperature state during
this phase (see discussion by Hoare \& Franco 2007).

Finally we note the recent 43 GHz observations of Orion-I by Reid et
al.\ (2007) which revealed an equatorial geometry for the wind. In
their case the geometry is unambiguous as the SiO maser emission
defines the outflow direction. They did not consider a wind origin for
this emission and interpreted their data in terms of an ionized
accretion disk instead. It is interesting to note the geometry of the
43-GHz emission from Orion-I is remarkably similar to that of
S\,140-IRS1, showing a twist or `kink' in the contours (Fig.\
\ref{s140}).

\section{Conclusion}

We have presented high-resolution observations made with the VLA in
its A configuration at frequencies between 5 and 43 GHz of a sample of
five massive young stellar objects (YSOs). The resolution varied from
0.04 arcsec (at 43 GHz) to 0.5 arcsec (at 5 GHz), corresponding to a
linear resolution between 17 and 85 AU.

We resolved elongated emission at 43 GHz in S106-IR and S140-IRS1
confirming the equatorial wind nature of the emission from these two
sources. This is further emphasized by a MERLIN 23-GHz observation of
S106-IR which reveals a 44$\times$18 AU$^2$ size for the radio source
perpendicular to the large-scale outflow. The other sources were also
partially resolved but were more difficult to relate to any bipolar
outflow activity due to either being too evolved or within too complex
an environment.

The spectral indices we derived are consistent with those for ionized
stellar winds, not \HII\ regions, with values ranging from +0.2 to
+0.7. The sizes of the sources also scaled with frequency as expected
for a stellar wind. We applied the procedure developed by Becker \&
White (1982) and directly model the visibility data in terms of an
isothermal, spherical wind. Reasonable fits were obtained for some of
the data. The derived wind temperatures of 3000--6000\,K are somewhat
lower than would be expected if the wind is heated by a main sequence
star. If the wind is heated mostly by an accretion disc or by a low
effective temperature star swollen by ongoing accretion this could
explain the low wind temperatures. The observed brightness
temperatures were also much lower than 10\,000\,K and indicated that
there is likely to be significant beam dilution so the wind may also
be clumpy or mixtures of optically thick and thin gas within the beam.

Further studies are needed to make use of the increased sensitivity,
resolution and spatial frequency coverage of the VLA, to probe the
ionized gas as close to the YSO as possible. However, in order to
maintain sensitivity to larger spatial scales, it is important to
obtain long (i.e.\ non-snapshot) tracks in multiple configurations,
and at multiple wavelengths so that a more accurate picture can be
derived for the ionized gas on circumstellar scales up to the size of
molecular outflows. More sophisticated non-spherical models of the
radio emission from equatorial winds and jets with temperature
structure will be needed to compare with such data.

\section*{ACKNOWLEDGMENTS}

We would like to thank Chris Carilli at the AOC, Socorro, for help
with the initial data reduction and Helmut Wiesemeyer for help with
reducing the PdBI data. The referee is thanked for comments which
helped clarify aspects of this work. This work was supported by grants
from PPARC to the University of Leeds. The National Radio Astronomy
Observatory is operated by Associated Universities, Inc., under
cooperative agreement with the National Science Foundation.

\bsp


\begin{thebibliography}{}
\bibitem{altenhoff} Altenhoff W.J., Thum C., Wendker H.J., 1994, A\&A,
  281, 161
\bibitem{alvarez04} Alvarez C., Hoare M.G., Glindemann A., Richichi
  A., 2004, A\&A, 427, 505
\bibitem{bsp83} Bally J., Snell R.L., Predmore R., 1983, ApJ, ApJ,
  272, 154
\bibitem{bally98} Bally J., Yu K.C., Rayner J., Zinnecker H., 1998,
  AJ, 116, 1868
\bibitem{bw82} Becker R.H., White R.L., 1982, ApJ, 262, 657
\bibitem{bw88} Becker R.H., White R.L., 1988, ApJ, 324, 893
\bibitem{b84} Bieging J.H., 1984, ApJ, 286, 591
\bibitem{bcs84} Bieging J.H., Cohen M., Schwartz P.R., 1984, ApJ, 282,
  699 
\bibitem{bik03} Bik A., Lenorzer A., Kaper L., Comer\'on F., Waters
  L.B.F.M., de Koter A., Hanson M.M., 2003, A\&A, 404, 249
\bibitem{bunn} Bunn J.C., Hoare M.G., Drew J.E., 1995, MNRAS, 272, 346
\bibitem{burrows} Burrows C.J., Stapelfeldt K.R., Watson A.M., Krist
  J.E., Ballester G.E., Clarke J.T., Crisp D., Gallagher J.S., III,
  et al.\
  1996, ApJ, 473, 437
\bibitem{ch} Carilli C., Holdaway M.A., 1999, Rad Sci, 34, 817
\bibitem{chandler95} Chandler C.J., Carlstrom J.E., Scoville N.Z.,
  1995, ApJ, 446, 793
\bibitem{davis98b} Davis C.J., Moriarty-Schieven G.H., Eisl\"offel J.,
  Hoare M.G., Ray T.P., 1998, AJ, 115, 1118
\bibitem{drew1989} Drew J.E., 1989, ApJS, 71, 267
\bibitem{dbh93} Drew J.E., Bunn J.C., Hoare M.G., 1993, MNRAS, 265, 12
\bibitem{dps98} Drew J.E., Proga D., Stone J.M., 1998, MNRAS, 296, L6
\bibitem{f85} Felli M., Simon M., Fischer J., Hamann F., 1985, A\&A,
  145, 305
\bibitem{g04} Giardino G., Favata F., Micela G., 2004, A\&A, 424, 965
\bibitem{ghlw03} Gibb A.G., Hoare M.G., Little L.T., Wright M.C.H.,
  2003, MNRAS, 339, 1011
\bibitem{ghmw} Gibb A.G., Hoare M.G., Mundy L.G., Wyrowski F., 2004,
  in Burton M.G., Jayawardhana R., \& Bourke T.L., eds, IAU Symposium
  221: Star Formation at High Angular Resolution. Astron. Soc. Pac.,
  San Francisco, p. 425
\bibitem{h04} Herbig G.H., Andrews S.M., Dahm S.E., 2004, AJ, 128,
  1233
\bibitem{h02} Hoare M.G., 2002, in Crowther P.A., ed., The Earliest
  Phases of Massive Star Birth. Astron. Soc. Pac., San Francisco,
  p. 137
\bibitem{h06} Hoare M.G., 2006, ApJ, 649, 856
\bibitem{h94} Hoare M.G., Drew J.E., Muxlow T.B., Davis R.J., 1994,
  ApJ, 421, L51
\bibitem{hf06} Hoare M.G., Franco J., 2007, in Hartquist T.W., Falle
  S.A.E., Pittard J.M., Diffuse Matter from Star Forming Regions to
  Active Galaxies -- A Volume Honouring John Dyson. Springer,
  Dordrecht, p. 61
\bibitem{hg95} Hoare M.G., Garrington S.T., 1995, ApJ, 449, 874
\bibitem{hgr96} Hoare M.G., Glindemann A., Richichi A., 1996, in
  K\"aufl H.U., Siebenmorgen R., eds., The Role of Dust in the
  Formation of Stars. Springer Verlag, Berlin, p. 35
\bibitem{hm96} Hoare M.G., Muxlow T.B., 1996, in Taylor A.R., Paredes
  J.M., eds., Radio Emission from the Stars and the Sun. Astron.
  Soc. Pacific, San Francisco, p. 47
\bibitem{hf87} H\"oflich P., Wehrse R., 1987, A\&A, 185, 107
\bibitem{hunter94} Hunter T.R., Taylor G.B., Felli M., Tofani G.,
  1994, A\&A, 284, 215
\bibitem{keto03} Keto E., 2003, ApJ, 599, 1206
\bibitem{k94} Kurtz S.E., Churchwell E., Wood D.O.S., 1994, ApJS, 91,
  659
\bibitem{l04} Lenorzer A., Bik A., de Koter A., Kurtz S.E., Waters
  L.B.F.M., Kaper L., Jones C.E., Geballe T.R., 2004, A\&A, 414, 245
\bibitem{marti93} Mart\'{\i} J., Rodr\'{\i}guez L.F., Reipurth B.,
  1993, ApJ, 416, 208
\bibitem{martin96} Martin S.C., 1996, ApJ, 473, 1051
\bibitem{menten} Menten K.M., van der Tak F.F.S., 2004, A\&A, 414, 289
\bibitem{mww} Minchin N.R., Ward-Thompson D., White G.J., 1993a, A\&A,
  298, 894
\bibitem{mwp} Minchin N.R., White G.J., Padman R., 1993b, A\&A, 277, 595
\bibitem{m83} Moran J.M., Garay G., Reid M.J., Genzel R., Wright
  M.C.H., Plambeck R.L., 1983, ApJ, 271, L31
\bibitem{n94} Nisini B., Smith H.A., Fischer J., Geballe T.R., 1994,
  A\&A, 290, 463
\bibitem{ouyed2003} Ouyed R., Clarke D.A., Pudritz R., 2003, ApJ, 582, 292
\bibitem{pf75} Panagia N., Felli M., 1975, A\&A, 39, 1
\bibitem{reipurth2002} Reipurth B., Heathcote S., Morse J., Hartigan
  P., Bally J., 2002, AJ, 123, 362
\bibitem{r07} Reid M.J., Menten K.M., Greenhill L.J., Chandler C.J.,
  2007, ApJ, accepted (arXiv:astro-ph/0704.2309)
\bibitem{r86} Reynolds S.P., 1986, ApJ, 304, 713
\bibitem{ry97} Richling S., Yorke H.W., 1997, A\&A, 327, 317
\bibitem{r94} Rodr\'{\i}guez L.F., Garay G., Curiel S., Ram\'{\i}rez
  S., Torrelles J.M., G\'omez Y., Vel\'azquez A., 1994, ApJ, 430, L65
\bibitem{rgr03} Rodr\'{\i}guez L.F., G\'omez Y., Reipurth B., 2003,
  ApJ, 598, 1100
\bibitem{r2005} Rodr\'{\i}guez L.F., Paveda A., Lizano S., Allen C.,
  2005, ApJ, 637, L65
\bibitem{miriad} Sault R.J., Teuben P.J., Wright M.C.H., 1995, in Shaw
  R.A., Payne H.E., Hayes J.J.E., eds., Astronomical Data Analysis
  Software and Systems IV. Astron. Soc. Pacific, San Francisco, p. 433
\bibitem{s00} Schertl D., Balega Y., Hannemann T., Hofmann K.-H,
  Preibisch T., Weigelt G., 2000, A\&A, 361, L29
\bibitem{sb82} Schmid-Burgk J., 1982, A\&A, 108, 169
\bibitem{s89} Schwartz P.R., 1989, ApJ, 338, L25
\bibitem{s01} Shepherd D.S., 2001, ApJ, 546, 345
\bibitem{sts03} Shepherd D.S., Testi L., Stark D.P., 2003, ApJ, 584, 882
\bibitem{skt04} Shepherd D.S., Kurtz S.E., Testi L., 2004, ApJ, 601, 952
\bibitem{s05} Sim S.A., Drew J.E., Long K.S., 2005, MNRAS, 363, 615
\bibitem{simon83} Simon M., Felli M., Cassar L., Fischer J., Massi M.,
  1983, ApJ, 266, 623
\bibitem{smith01} Smith N., Jones T.J., Gehrz R.D., Klebe D.,
  Creech-Eakman M.J., 2001, AJ, 121, 984
\bibitem{t95} Tofani G., Felli M., Taylor G.B., Hunter T.R., 1995,
  A\&AS, 112, 299
\bibitem{t97} Torrelles J.M., G\'omez J.F., Rodr\'{\i}guez L.F., Ho
  P.T.P., Curiel S., V\'azquez R., 1997, ApJ, 489, 744
\bibitem{t1} Trinidad M.A., Curiel S., Cant\'o J., D'Alessio P.,
  Rodr\'{\i}guez L.F., Torrelles J.M., G\'omez J.F., Patel N., Ho
  P.T.P., 2003, ApJ, 589, 386
\bibitem{tuthill02} Tuthill P.G., Monnier J.D., Danchi W.C., Hale
  D.D.S., Townes C.H., 2002, ApJ, 577, 826
\bibitem{vdt1} van der Tak F.F.S., Menten K.M., 2005, A\&A, 437, 947
\bibitem{vdt2} van der Tak F.F.S., Tuthill P.G., Danchi W.C., 2005,
  A\&A, 431, 993
\bibitem{vdt3} van der Tak F.F.S., van Dishoeck E.F., Evans N.J., II,
  Blake G.A., 2000, ApJ, 537, 283
\bibitem{ww94} Wilner D.J., Welch W.J., 1994, ApJ, 427, 898
\bibitem{w99} Wilner D.J., Reid M.J., Menten K.M., 1999, ApJ, 513, 775
\bibitem{wb75} Wright A.E., Barlow M.J., 1975, MNRAS, 170, 41

\end{thebibliography}
\end{document}